\documentclass[preprint]{aastex}
\usepackage{geometry}
\usepackage{graphicx}
\usepackage{epsfig}

\begin{document}
\title{Direct evidence for a supernova interacting with a large amount of hydrogen-free circum-stellar material.}
\author{Sagi Ben-Ami}
\affil{Department of Particle Physics and Astrophysics, The Weizmann Institute of Science, Rehovot 76100, Israel.}
\email{sagi.ben-ami@weizmann.ac.il}   
\author{Avishay Gal-Yam}
\affil{Department of Particle Physics and Astrophysics, The Weizmann Institute of Science, Rehovot 76100, Israel.}   
\affil{Kimmel Investigator.}
\author{Paolo A. Mazzali}
\affil{Astrophysics Research Institute, Liverpool John Moores University. Liverpool L3 5RF, UK.}
\affil{Max-Planck-Institut fur Astrophysik, Karl-Schwarzschildstr. 1, D-85748 Garching, Germany.}
\affil{INAF - Osservatorio Astronomico, vicolo dell'Osservatorio, 5, I-35122 Padova, Italy.}
\author{Orly Gnat} 
\affil{Racah Institute of Physics, The Hebrew University, 91904 Jerusalem, Israel.}
\author{Maryam Modjaz}
\affil{Center for Cosmology and Particle Physics, Department of Physics, New York University, 4 Washington Place, room 529, New York, NY 10003, USA.}
\author{Itay Rabinak}
\affil{Department of Particle Physics and Astrophysics, The Weizmann Institute of Science, Rehovot 76100, Israel.} 
\author{Mark Sullivan}
\affil{School of Physics \& Astronomy, University of Southampton, Highfield, Southampton, SO17 1BJ, U.K.}
\author{Lars Bildsten}
 \affil{Kavli Institute for Theoretical Physics and Department of Physics Kohn Hall, University of California, Santa Barbara, CA 93106, USA.}
\author{ Dovi Poznanski}
\affil{School of Physics and Astronomy, Tel-Aviv University, Tel Aviv, 69978 Israel.}
\author{Ofer Yaron, Iair Arcavi}
\affil{Department of Particle Physics and Astrophysics, The Weizmann Institute of Science, Rehovot 76100, Israel.}   
\author{Joshua S. Bloom}
\affil{Department of Astronomy, University of California, Berkeley, CA 94720-3411, USA.}
\affil{Physics Division, Lawrence Berkeley National Laboratory, Berkeley, CA 94720, USA. }
\author{Assaf Horesh}
\affil{Cahill Center for Astrophysics, California Institute of Technology, Pasadena, CA, 91125, USA.}
\author{Mansi M. Kasliwal}
\affil{Observatories of the Carnegie Institution for Science, 813 Santa Barbara St., Pasadena, CA 91101, USA.}
\author{Shrinivas R. Kulkarni}
\affil{Cahill Center for Astrophysics, California Institute of Technology, Pasadena, CA, 91125, USA.}
\author{Peter E. Nugent}
\affil{Department of Astronomy, University of California, Berkeley, CA 94720-3411, USA.}
\affil{Physics Division, Lawrence Berkeley National Laboratory, Berkeley, CA 94720, USA. }
\author{Eran O. Ofek}
\affil{Department of Particle Physics and Astrophysics, The Weizmann Institute of Science, Rehovot 76100, Israel.}   
\author{Daniel Perley} 
\affil{Cahill Center for Astrophysics, California Institute of Technology, Pasadena, CA, 91125, USA.}
\author{Robert Quimby}
\affil{Kavli IPMU, University of Tokyo, 5-1-5 Kashiwanoha, Kashiwa-shi, Chiba, 277-8583, Japan.}
\author{Dong Xu}
\affil{Dark Cosmology Centre, Niels Bohr Institute, University of Copenhagen, Juliane Maries Vej 30, 2100, Copenhagen, Denmark. }
\begin{abstract}
We present our observations of SN 2010mb, a Type Ic SN lacking spectroscopic signatures of H and He.
SN 2010mb has a slowly-declining light curve ($\sim600\,$days) that cannot be powered by $^{56}$Ni/$^{56}$Co radioactivity, the common energy source for Type Ic SNe. 
We detect signatures of interaction with hydrogen-free CSM including a blue quasi-continuum and , uniquely, narrow oxygen emission lines that require high densities ($\sim10^9$cm$^{-3}$). 
From the observed spectra and light curve we estimate that the amount of material involved in the interaction was $\sim3$M$_{\odot}$. 
Our observations are in agreement with models of pulsational pair-instability SNe described in the literature.
\end{abstract}
\section{Introduction}
A massive star with an initial mass of $>8$M$_{\odot}$ ends its life in an explosion that destroys the star, 
leaving a neutron star (NS) or a black hole (BH) as a remnant \citep{Heger2003}. This explosion is triggered by iron photo-disintegration and loss of internal energy causing 
the star to undergo a gravitational core-collapse supernova (CC SN; e.g., Woosley \& Janka 2005).\\ 
An extremely massive star with an initial mass above $\sim140$M$_{\odot}$ will avoid reaching photo-disintegration, and instead will go through a phase of 
electron-positron generation in its core at an earlier stage of its evolution \citep{Barkat1967,Rakavy1968,Heger2003,Waldman2008,Chatz2012}. The pair production will render 
the star unstable, causing the star to end its life in an explosion that ejects the entire mass of the star and leaves no remnant at all, a pair instability supernova (PISN).
SN 2007bi and PTF10nmn are candidate examples of PISNe, both with an estimated ejecta mass $\sim100$M$_{\odot}$\,  (Gal-Yam et al. 2009; Gal-Yam 2012; Yaron et al. 2013 in prep). In both cases the 
explosion is luminous, with the light curve of the event as predicted by the radioactive decay chain $^{56}$Ni$\longrightarrow^{56}$Co$\longrightarrow^{56}$Fe, and with 
spectra lacking hydrogen lines, consistent with recent stellar evolution model of stars with initial mass $>140$M$_{\odot}$ \citep{Yusof2013}.\\
At the mass range in between CC SNe and PISNe, stars will enter a phase of electron-positron generation in the core as well, but the resulting explosion is insufficiently energetic 
to unbind the star \citep{Heger2002,Heger2003,Woosley2007,Waldman2008,Chatz2012}. The star will eject matter in a series of eruptions, a pulsation pair instability event (PPI), thus reducing the core mass
until it reaches hydrostatic equilibrium and returns to the normal evolution track of massive stars, ending its life in a CC SN. The matter ejected may be composed mainly of 
He, C, and O, if the star lost most of its hydrogen envelope earlier \citep{Chatz2012}. Interaction between subsequent shells, or between the ejecta of an ultimate CC SN 
explosion and a shell ejected at an earlier stage, are possible signatures of these events. So far, only indirect evidence for a PPI event has been discussed, e.g., for 
SN 2006gy and SN 2006jc \citep{Woosley2007,Pastorello2008a,Chugai2009}.\\
Here, we describe SN 2010mb (PTF10iue), a SN lacking signatures of either hydrogen or helium (Type Ic), with long-lasting emission powered by interaction of the SN ejecta with a large mass of 
hydrogen-free CSM. Section $\S2$ describes our observations; and section $\S3$ presents our results and analysis. In section $\S4$ we discuss possible scenarios, with an 
emphasis on the PPI option. Conclusions are given in section $\S5$.                      
\section{Observations}
\subsection{Discovery}
On April 10 2010 UT the Palomar Transient Factory (PTF;  Law et al. 2009, Rau et al. 2009) detected SN 2010mb at R.A. $ =16$h$00$m$23$s$.103$  and Decl.$ =  37^{\circ}44'56''.98$ using the 
CFH12K survey camera mounted on the 48'' Oschin Schmidt telescope at the Palomar observatory (P48). 
Analysis of previous images showed the SN was visible on March 18 2010, at a magnitude of $22.0\pm0.5$ in the $r$-band.
The object is located on the edge of the galaxy \textit{SDSS J160023.23+374454.8} at a redshift z$=0.1325$. Fig. 1 shows detection, reference, and subtracted images of the SN.\\
\begin{figure}[h!p!]
\centerline{
\begin{tabular}{cc}
\scalebox{0.36}{\includegraphics{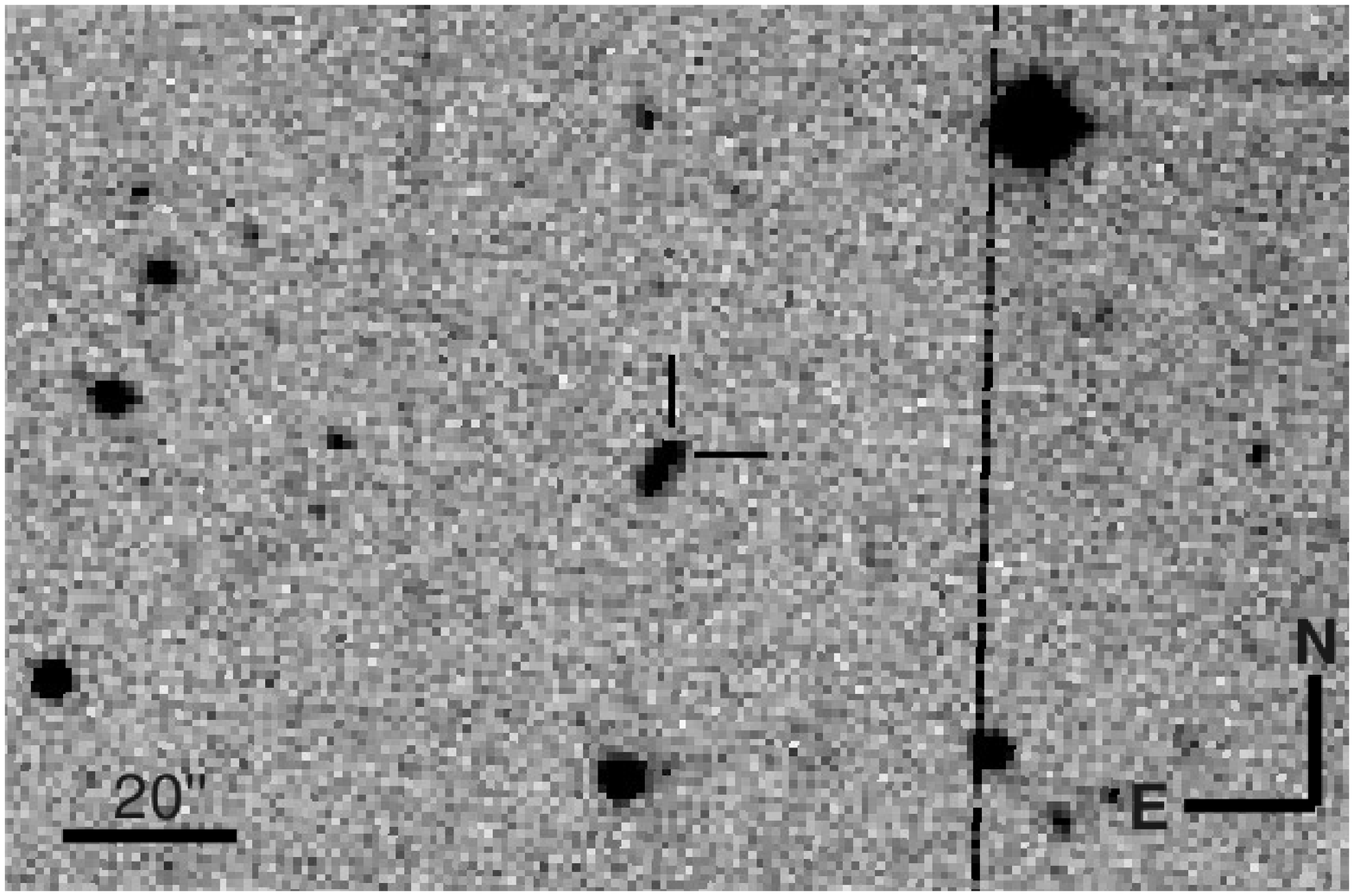}} &
\scalebox{2.4}{\includegraphics{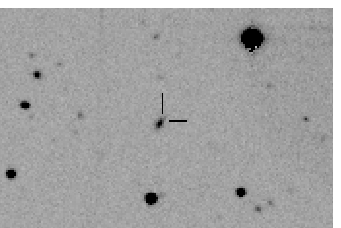}} \\
\scalebox{2.4}{\includegraphics{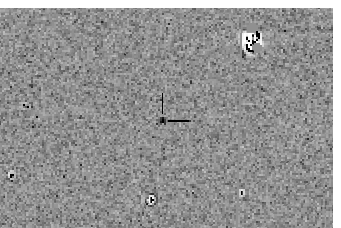}} &
\scalebox{0.3}{\includegraphics{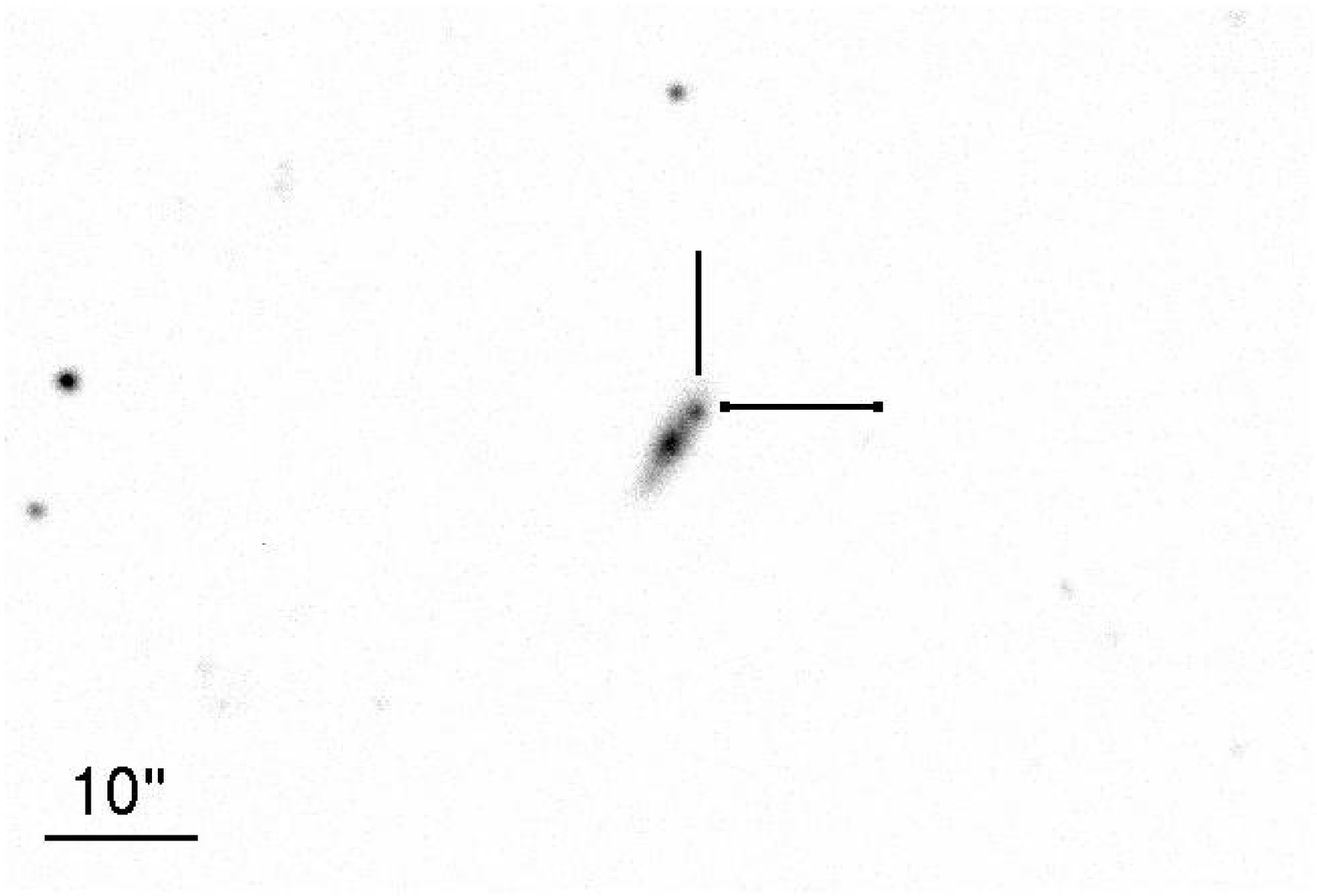}} \\
\end{tabular}}
 \renewcommand{\thefigure}{1}
\caption{ \small 
          \textbf{Top Left:} An image of SN 2010mb taken with the P48 On UT 2010 June 10.
          \textbf{Top Right:} Reference image for SN 2010mb.
          \textbf{Bottom Left:} The SN 2010mb image after reference subtraction. Residuals from host galaxy contamination are negligible, as is apparent in the image. 
          \textbf{Bottom Right:} A zoomed image of SN 2010mb obtained on March 4 2011 with LRIS mounted on the Keck-I 10m telescope. The offset of the SN from the host center 
                                 ($0.1''$ W and $2.2''$ N) is clearly seen.
    }
\end{figure} 
Prior to March 18 2010, the galaxy was imaged $10$ times by the PTF survey between May 18 2009 and August 21 2009, with no evidence for the SN or any activity in its vicinity.\\
We classified SN 2010mb as a Type Ic SN based on a spectrum lacking signatures of neither hydrogen nor helium taken on June 8 2010 (Fig. 3 and Fig. 5).
\subsection{Photometry}
SN 2010mb, discovered near the P48 detection limit, was intermittently detected during the first 50 days after its discovery\footnote{We define a detection as a $5\sigma$ 
signal above the zero point photon count. In cases where the object did not pass this criterion, April 30 and May 12 2010, the signals were $4\sigma$ and $2.6\sigma$ above 
the zero point photon count respectively.}, until it became continuously visible on May 31 2010 at an apparent magnitude of $20.85\pm0.10$ in the r band.
Photometry of SN 2010mb was obtained by the P48, the GRB camera \citep{Cenko2006} mounted on the Palomar 60'' telescope (P60), the Large Format Camera mounted on the 
Palomar 200'' Hale telescope (P200), and the Low Resolution Imaging Spectrograph (LRIS) mounted on the 10m Keck-I telescope  \citep{Oke1995}. Data were reduced using the 
MKDIFFLC photometry routine  \citep{GalYam2004,GalYam2008}, except for P48 data reduced using PSF photometry on image subtractions (e.g., Ofek et al. 2013). 
We adopt a distance modulus of $38.88\,$mag, corresponding to a distance of $\approx 599$\,Mpc, a Galactic extinction correction of $0.04\,$mag at r-band, and a reddening 
$E(B-V) = 0.015\,$mag, taken from NED\footnote{NASA/IPAC Extragalactic Database (NED) is operated by the Jet Propulsion Laboratory, California Institute of Technology, 
under contract with the National Aeronautics and Space Administration.}. Photometric results are given in Table 1 and plotted in Fig. 2\,. Possible host extinction is not corrected for. 
The lack of strong ISM absorption and the blue SED measured at late times (see $\S3$) argue against a substantial host extinction.\\
The light curve displayed a long plateau in $r$-band lasting $\sim180$ days, followed by a slow decline at a rate of $0.004$ magnitudes per day, while the $g$ and B-band 
light curves are slowly rising during the first $\sim250$ days (Fig. 2). The total amount of energy emitted in the r-band over a period of $\sim600$\,days is 
$\sim10^{50}$\,ergs (an average flux of $5.7\times10^{-14}$\,erg/s/cm$^2$). Assuming a bolometric correction of $10\%$ at early times and $50\%$ at later times based on
our spectral decomposition (see $\S3.2$), we estimate a total energy emission of $3.5\pm0.3\cdot10^{50}\,$erg over a period of $\sim600$\,days. 
\renewcommand{\arraystretch}{0.5}
\begin{table}[h!p!]
\centering
\tiny
\begin{tabular}{|c|c|c|c|c|c|}
  \hline
  Date (MJD) & Instrument & Exposure time [sec] & Apparent Magnitude [mag] & Magnitude Error & Filter \\ \hline\hline
  55273 & P48 & $2\times200$ & 22.2  & 0.54 & r \\
  55296 & P48 & $2\times200$ & 21.47 & 0.2  & r \\
  55301 & P48 & $2\times200$ & 20.93 & 0.16 & r \\
  55322 & P48 & $2\times200$ & 21.07 & 0.16 & r \\
  55336 & P48 & $2\times200$ & 20.81 & 0.19 & r \\ 
  55347 & P48 & $2\times200$ & 20.85 & 0.12 & r \\
  55352 & P48 & $2\times200$ & 20.96 & 0.13 & r \\
        & P60 & $360$ 	 	   & 20.76 & 0.18 & i \\	 
  55357 & P48 & $2\times200$ & 20.9  & 0.12 & r \\
  55358 & P48 & $2\times200$ & 20.8  & 0.15 & r \\
  55360 & P48 & $2\times200$ & 20.7  & 0.15 & r \\
  55365 & P48 & $2\times200$ & 20.75 & 0.13 & r \\
  55370 & P48 & $2\times200$ & 20.75 & 0.2  & r \\
  55375 & P48 & $2\times200$ & 20.35 & 0.15 & r \\
  55382 & P48 & $2\times200$ & 20.85 & 0.14 & r \\  
  55384 & P60 & $360$ & 21.0 & 0.1  & r \\ 
        & P60 & $240$ & 20.0 & 0.16 & B \\
        & P60 & $240$ & 22.0 & 0.2  & g \\
        & P60 & $240$ & 21.0 & 0.12  & i \\ 
  55387 & P60 & $180$ & 20.99& 0.2  & r \\
        & P60 & $360$ & 20.07 & 0.16 & B \\
  55396 & P60 & $180$ & 20.73 & 0.17 & r \\
        & P60 & $360$ & 20.63 & 0.16 & i \\ 
  55397 & P60 & $180$ & 20.65 & 0.2  & r \\
        & P60 & $360$ & 20.51 & 0.2  & B \\
  55406 & P60 & $360$ & 20.9  & 0.13 & r \\ 
        & P60 & $360$ & 20.67 & 0.19 & i \\ 
  55409 & P60 & $360$ & 20.04 & 0.16 & B \\
  55411 & P60 & $360$ & 21.02 & 0.12 & r \\
        & P60 & $360$ & 19.8 & 0.2  & B \\
        & P60 & $360$ & 20.64 & 0.12  & i \\
  55412 & P60 & $360$ & 21.6 & 0.2  & g \\            
  55415 & P60 & $360$ & 21.02 & 0.15 & r \\
  55417 & P60 & $360$ & 20.61 & 0.15 & i \\ 
  55422 & P60 & $180$ & 20.95 & 0.13 & r \\ 
        & P60 & $360$ & 19.9 & 0.16 & B \\
        & P60 & $360$ & 21.8 & 0.2  & g \\ 
        & P60 & $360$ & 20.98 & 0.16  & i \\ 
  55477 & LFC & $900$ & 20.78 & 0.15 & r \\                 
  55499 & LFC & $1200$& 21.71 & 0.14 & r \\
  55539 & P60 & $360$ & 20.1 & 0.2  & B \\
  55545 & P60 & $360$ & 20.0 & 0.15 & B \\
  55563 & P60 & $360$ & 20.02 & 0.14 & B \\
        & P60 & $360$ & 20.83 & 0.13  & g \\ 
  55573 & P60 & $360$ & 19.8  & 0.14 & B \\ \
        & LFC & $2400$& 22.53 & 0.14 & r \\
  55624 & LRIS& $120$ & 22.6  & 0.14 & r \\ 
  55677 & LRIS& $540$ & 22.8  & 0.2 & r \\
  55744 & LRIS& $720$ & 22.84 & 0.15 & r \\
  55836 & LFC & $2800$& 23.12 & 0.14 & r \\
  55916 & LFC & $3000$& 24.1  & Limit & r \\
  55952 & LRIS& $600$ & 24.8  & Limit & r \\ \hline
\end{tabular}
 \renewcommand{\thetable}{1}
   \label{tab1}
\caption{\small SN 2010mb photometry. 
Digital data are available from the Weizmann Interactive Supernova Data Repository (WISeREP; Yaron \& Gal-Yam 2012; \textit{http://www.weizmann.ac.il/astrophysics/wiserep/}).}
\end{table} 
\begin{figure}[h!p!]
\centerline{
\scalebox{0.9}{\includegraphics{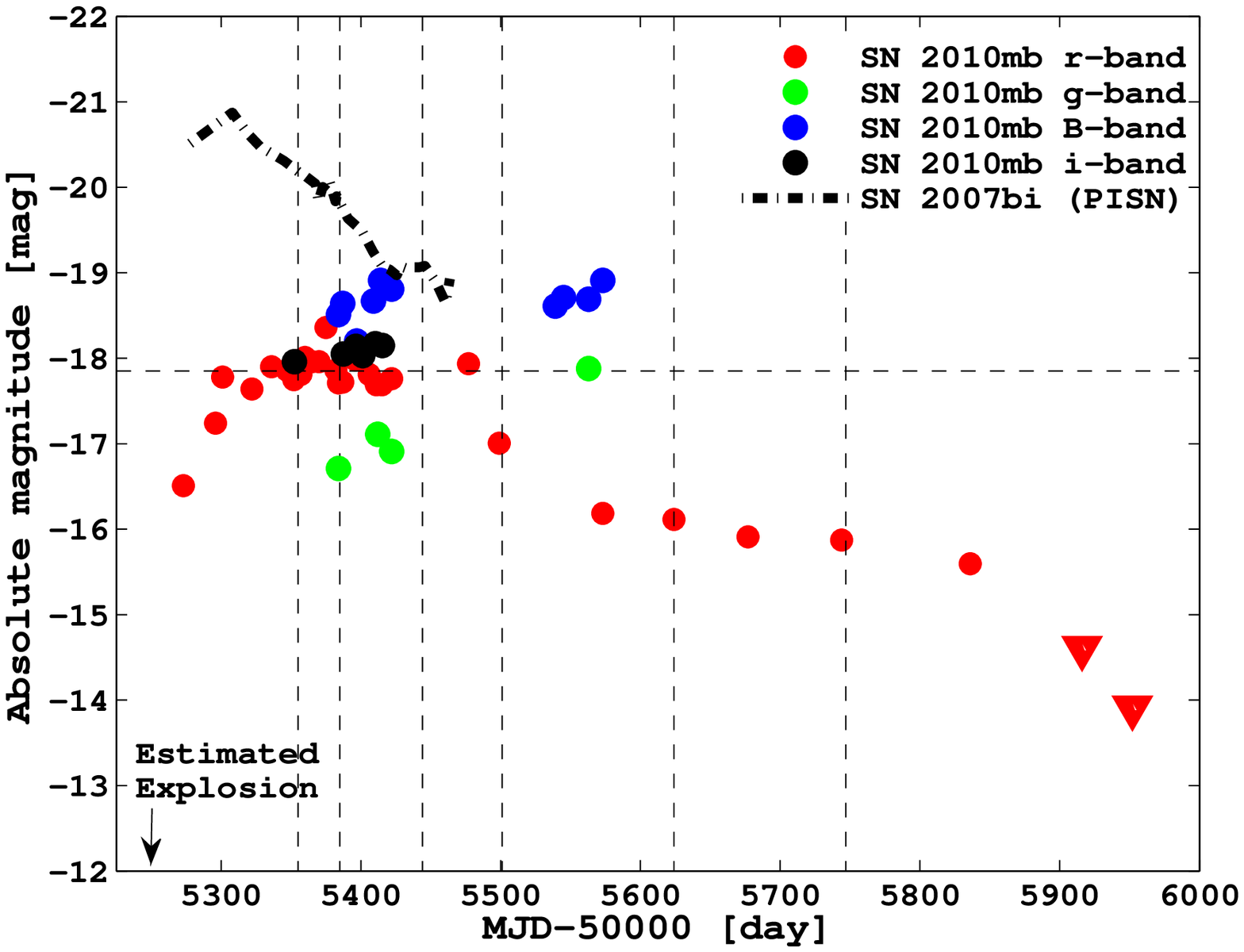}}
}
\caption{\small Photometry of SN 2010mb. The $r$-band light curve rises by $1.3$ magnitudes during the first $\sim25$ days, 
and then settles onto a plateau lasting $\sim180$ days at an absolute magnitude of $-17.8\,$mag (horizontal dashed line). Following the plateau, the LC drops by $0.9\,$mag in less than $23\,$days, 
followed by a period of slow decline at a rate of $0.004\,$mag/day, less than half the decline rate of  $^{56}$Ni/$^{56}$Co powered SNe, during the next $350$ days. 
Photometry in the $g$ and B bands showed a slow increase in flux during the first $250$ days of this event. The dashed vertical lines indicate a spectrum was taken on that 
day. The light curves of SLSN-R SN 2007bi (candidate pair instability event; Gal-Yam et al. 2009), and of an average Type Ic SN template \citep{Drout2011}, are shown for 
comparison. On December 21 2011, the SN was no longer visible in deep images taken with the LFC. Deeper images taken with LRIS on January 
26 2012 set a limiting absolute magnitude of $-13.91$ at that time (triangles).}
\end{figure}
\subsection{Spectroscopy}
Six spectra of SN 2010mb were obtained (Table 2, Fig. 3). 
Data were reduced using standard IRAF and IDL routines, and smoothed with a smooth 
box of three pixels. The observed spectra are deredshifted by $z=0.1325$. Spectra are calibrated to the gri/r P60/LFC photometry, Table 1\,. 
A spectrum of the host galaxy nucleus was taken on February 20 2012 using LRIS (Fig. 4).\\ 
\renewcommand{\arraystretch}{1.0}
\clearpage
\begin{table}[h!p!]
\centering
\tiny
\begin{tabular}{|c|c|c|c|c|c|c|c|}
  \hline
  Date           & Telescope   & Instrument                   &  Exposure Time[sec] & Grism/Grating [lpm]    &   Slit   & Resolution [\AA{}]            & Remarks  \\ \hline
  2010-06-08     &  Keck-I     &  LRIS                        &      600            &  400/400               &    1''   &   1.09/1.16                   & \\
  2010-07-08     &  Keck-I     &  LRIS                        &      750            &  400/400               &    1''   &   1.09/1.16                   & \\
  2010-09-05     &  Keck-I     &  LRIS                        &      750            &  400/400               &    1''   &   1.09/1.16                   & \\
  2010-11-07     &  Keck-I     &  LRIS                        &      750            &  400/400               &    1''   &   1.09/1.16                   & \\
  2011-03-04     &  Keck-I     &  LRIS                        &      900            &  400/400               &    1''   &   1.09/1.16                   & \\
  2011-07-05     &  Keck-II    &  DEIMOS                      &     1200            &  600                   &    0.7'' &   0.65                        & (Filter GG455) \\
  2012-02-20     &  Keck-I     &  LRIS                        &     1200            &  400/400               &    1''   &   1.09/1.16                   & Host-Galaxy spectrum \\ \hline
\end{tabular}
 \renewcommand{\thetable}{2}
  \label{tab2}
\centering
\caption{\small Spectroscopic observation log.}
\end{table}

\begin{figure}[h!p!]
\centerline{
 \scalebox{0.9}{\includegraphics{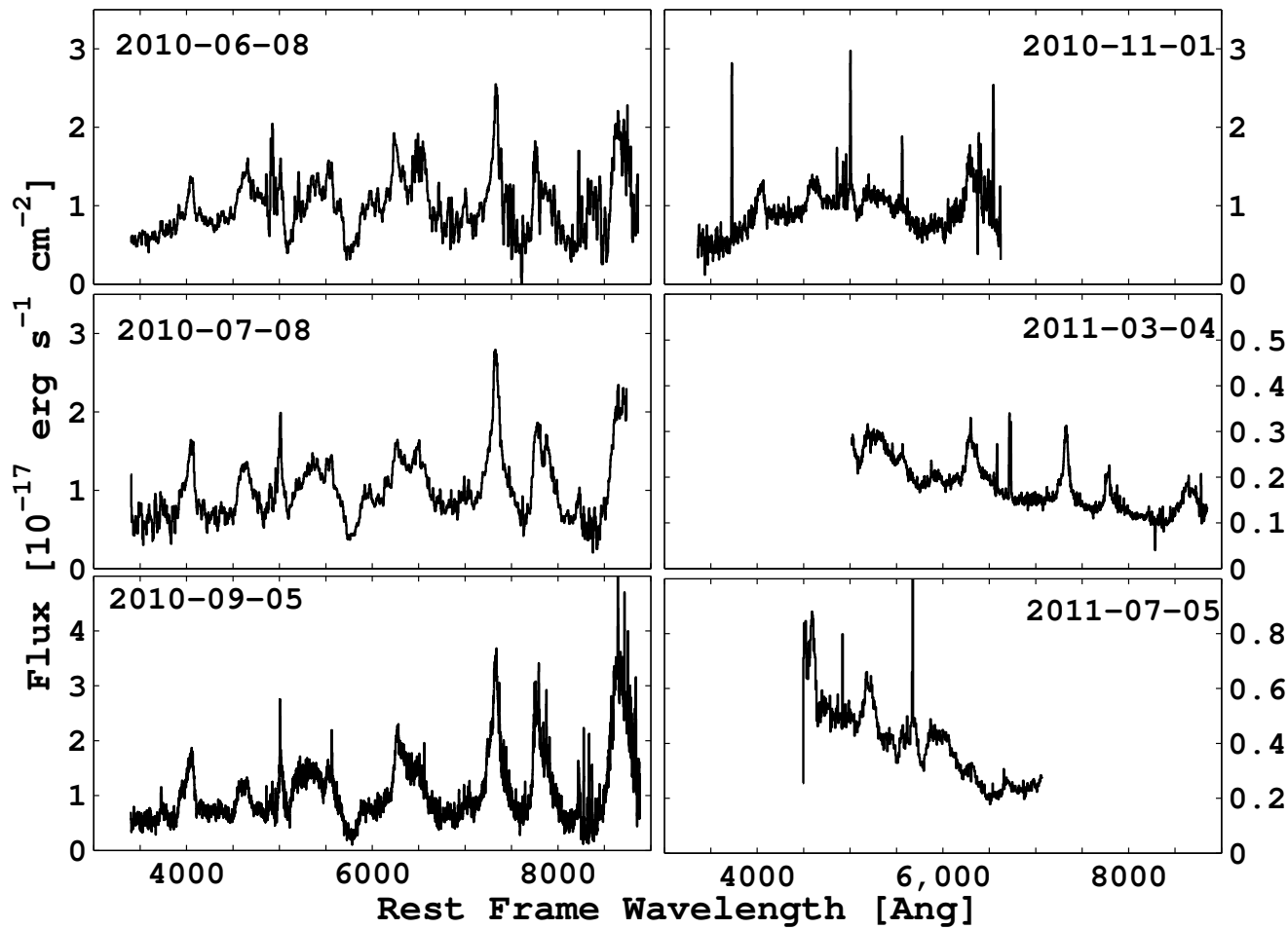}}
}
\renewcommand{\thefigure}{3}
\caption{\small SN 2010mb Spectra.
No signs of hydrogen lines (emission or absorption) or helium lines are seen, indicating a Type Ic SN. Late time spectra (March 4 2011 and July 5 2011) are dominated by a blue quasi-continuum component.
Spectra are available in digital form from the Weizmann Interactive Supernova Data Repository (WISeREP; Yaron \& Gal-Yam 2012; \textit{http://www.weizmann.ac.il/astrophysics/wiserep/}).
}
\end{figure}   
\begin{figure}[h!p!]
\centering
 \scalebox{0.9}{\includegraphics{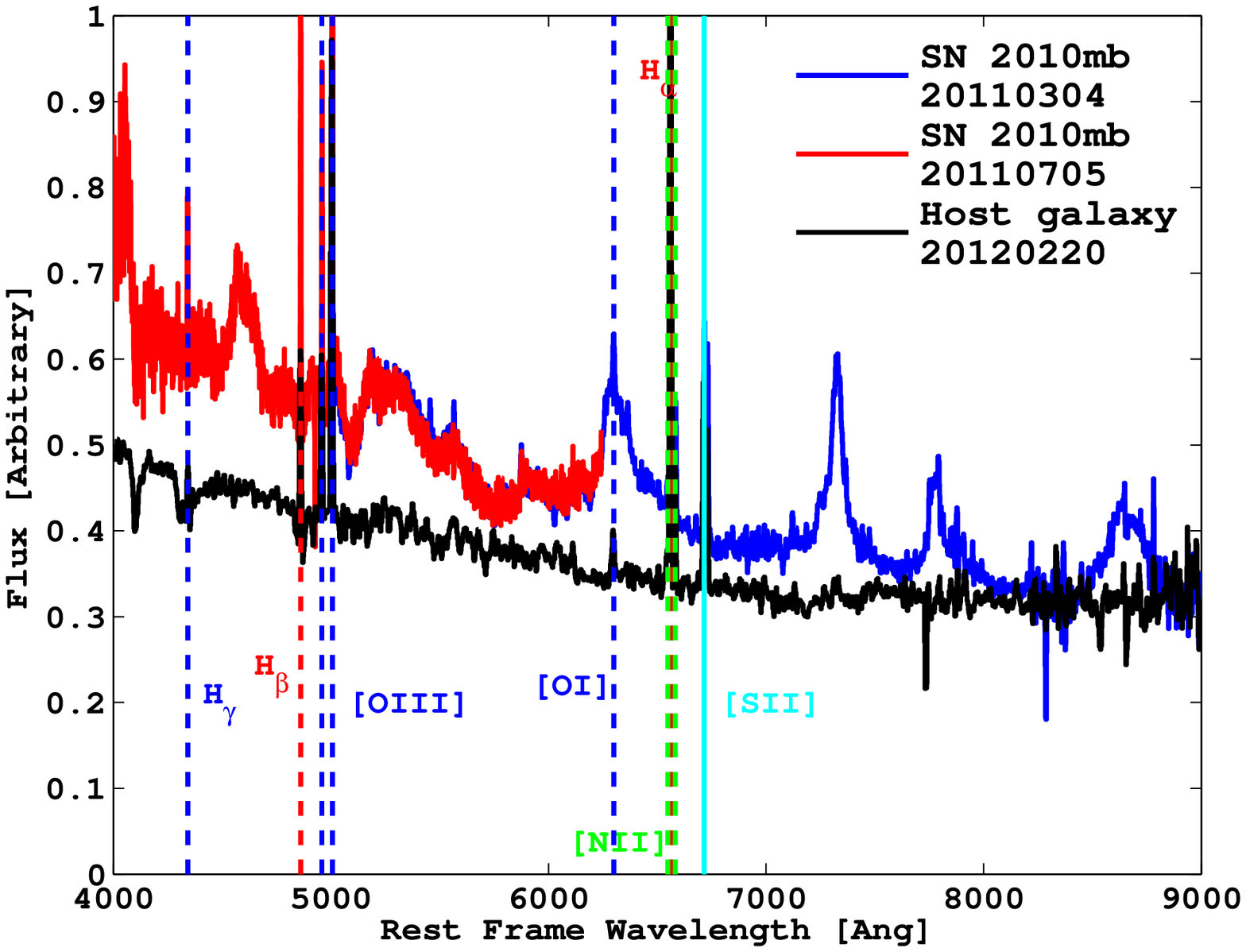}}
\renewcommand{\thefigure}{4}
  \label{fig4}
\caption{\small Host Galaxy nucleus Spectrum. 
The spectrum was taken on February 20 2012 using LRIS mounted on Keck-I $10$m telescope and is calibrated to r-band photometry. 
The SN has an offset of $0.1''$ W and $2.2''$ N from the host galaxy center.
The spectrum shows narrow Balmer emission lines filling slightly wider Balmer absorption features. This is typical to galaxies with a dominant population of older stars
($\sim1\,$Gyr) mixed with a population of younger hotter stars \citep{Dressler1983,Covino2006}.
Host-galaxy lines of H and O indicate a redshift of z$=0.1325$\,.
The galaxy has a metallicity of $0.5^{+0.42}_{-0.19}$Z$_{\odot}$, having used the scale of Pettini \& Pagel (2004) and a solar oxygen abundance value of 
$12+\log$(O/H)$=8.69$ \citep{Asplund2009}. The total star formation rate we measure using the H$\alpha$ luminosity (Kennicutt et al. 1998) is 
$\sim0.32\,$M$_{\odot}\,$yr$^{-1}\,$. Two spectra of SN 2010mb (2011-03-04, and 2011-07-05) are given for comparison.
Spectra are available in digital form from the Weizmann Interactive Supernova Data Repository (WISeREP; Yaron \& Gal-Yam 2012; \textit{http://www.weizmann.ac.il/astrophysics/wiserep/}).
}
\end{figure}    
\subsection{X-ray}
SN 2010mb was imaged with the Swift X-ray telescope  \citep{Burrows2005} on April 22 2011. Using an exposure of $5435$s we derive a $3\sigma$ upper 
limit of $5\times10^{-14}$erg/s/cm$^2$ for the $0.3-10$keV band assuming a photon index of 2.\\

\section{Analysis}
\subsection{Classification and Evolution}
The first three spectra of SN 2010mb (June 8, July 8, and September 5 2010) resemble a Type Ic SN at the transition between photospheric and nebular phase as late as 
$\sim190$\,days after discovery (i.e., photospheric features observed in the spectrum taken on September 5 2010, Fig. 3).
Automatic classification using Superfit \citep{Howell2005} of the spectra taken on June 8 and July 8 2010 suggests that the best match is to the late-time spectra of 
SN 1997ef and SN 1995F at 89 and 90 days after discovery/peak magnitude respectively (Mazzali et al. 2004; Matheson et al. 2001; Fig. 5), while automatic classification using SNID (Blondin \& Tonry 2011) suggests that the best match is to SN 2007gr at 54 days after peak magnitude (Hunter et al. 2009). This highlights the slow evolution of SN 2010mb, similar to SN 1997ef, and much slower than more normal SNe Ic like SN 2007gr
The spectra also resemble nebular spectra of SN 2007bi (Gal-Yam et al. 2009; Fig. 5). Since both SN 1997ef and SN 1995F showed a late transition from photospheric to nebular phase \cite{Mazzali2004}. Based on the resemblance between SN 2010mb spectrum taken on June 8 2010, and SN 1997ef spectrum taken $89\,$days after peak magnitude, and assuming a rise time of $20\,$days \citep{Mazzali2004}, we assume the SN 2010mb exploded on February 23 2010, though this number is highly uncertain.
Early and intermediate spectra (i.e June 8 until November 1 2010) are dominated by blended lines of CaII and [SII] ($3933$, $3968$, and $4068$, $4076$\AA), 
[MgI] and [SI] ($4562$, $4571$, $4507$, and $4589$\AA), [OIII] ($4959$ and $5007$\AA), FeII blended lines ($5200-5400$\AA) ,[OI] and FeII blended lines ($6300-6400$\AA), OI and [SI] ($7773$ and $7728$\AA), and CaII and [CI]  ($8498, 8542, 8662$ and $8727$\AA). The NaI D line ($5889$\AA), [SiI] line ($6527$\AA), and [CaII] line ($7291$, $7324$\AA) are clearly observed in the spectra as well (Line identification is based on the models described in $\S3.2$; Wavelengths are given in restframe), Fig. 5.  
Some of the lines seem to have an internal structure, an indication for a non-spherical geometry, Fig. 6. 
Late time spectra (March 4 2011 and July 5 2011, Fig. 3) are dominated by a blue quasi-continuum component, also reflected in $g$ and B-band photometry that is slowly rising 
during the first $\sim250$ days, a behaviour inconsistent with that of a purely hydrodynamic radioactive explosion that would show a monotonic decrease in temperature with 
time.\\
\begin{figure}[h!p!]
  \centerline{
    \scalebox{0.9}{\includegraphics{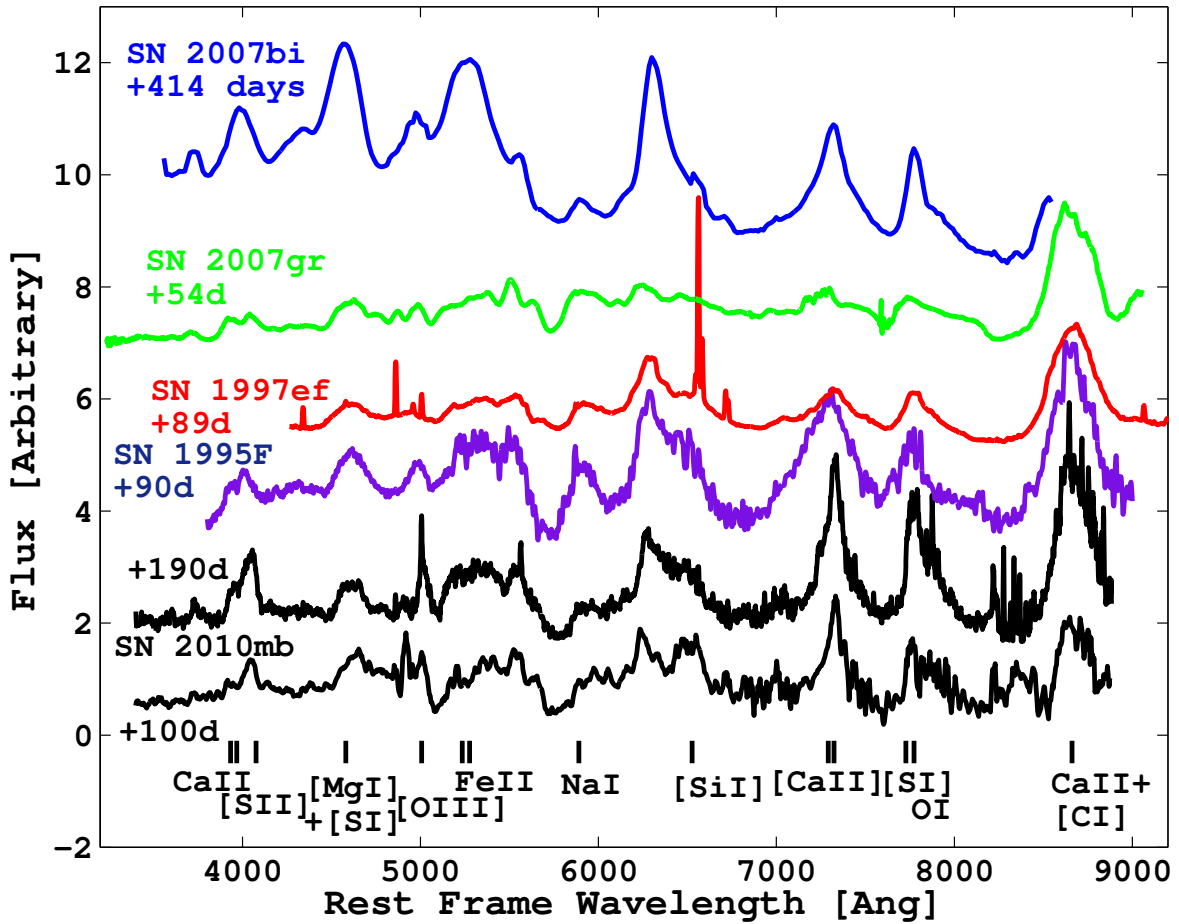}}
  }
   \renewcommand{\thefigure}{5}
  \caption{\small SN 2010mb Spectral Classification. 
           Early spectra of SN 2010mb resemble a Type Ic SN at the transition between photospheric and nebular phase as late as $\sim190$\,days after discovery (i.e. 
           some photospheric features still observed in the spectrum taken on September 5 2011, second spectrum from the bottom). 
           Spectra are dominated by blended lines of CaII and [SII] ($3933$, $3968$, and $4068$, $4076$\AA), 
           [MgI] and [SI] ($4562$, $4571$, $4507$, and $4589$\AA), [OIII] ($4959$ and $5997$\AA), FeII blended lines ($5200-5400$\AA) ,[OI] and FeII blended lines ($6300-6400$\AA), 
           OI  and [SI] ($7773$ and $7728$\AA), and CaII and [CI]  ($8498, 8542, 8662$ and $8727$\AA). The NaI D line ($5889$\AA), [SiI] line ($6527$\AA), and [CaII] line ($7291$, 
           $7324$\AA) are clearly observed in the spectra as well (Line identification is based on the models described in $\S3.2$; Wavelengths are given in restframe).  
           An automatic classification using Superfit \citep{Howell2005} finds the best match is to the late-time spectra of SN 1997ef and SN 1995F at $89$ and
           $90$\,days after discovery/peak magnitude respectively \citep{Mazzali2004,Matheson2001},
           while automatic classification using SNID (Blondin \& Tonry 2011) suggests that the best match is to SN 2007gr at 54 days after peak magnitude (Hunter et al. 2009).
           The spectra also resemble nebular spectra of SN 2007bi \citep{GalYam2009}.
          }
\end{figure}

\begin{figure}[h!p!]
  \centerline{
    \scalebox{0.85}{\includegraphics{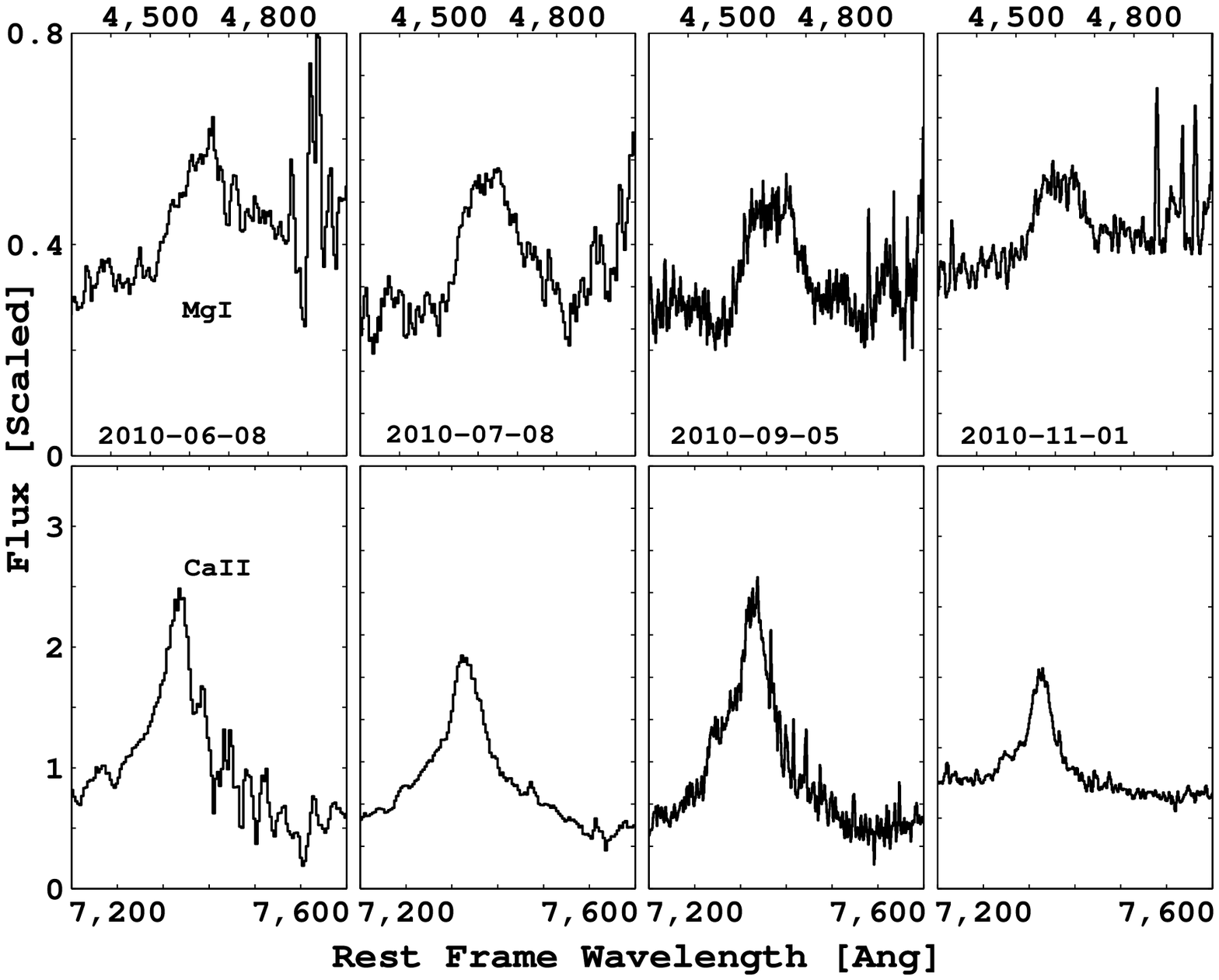}}
  }
   \renewcommand{\thefigure}{6}
  \caption{\small SN 2010mb line profiles. 
           Comparing the nebular emission lines from MgI (top panels) and CaII (bottom panels), we see a flat-topped profile in the former line, 
           usually interpreted as a signature of non-spherical geometry \cite{Maeda2008,Mazzali2005,Modjaz2008,Taubenberger2009}.
           The different line shapes of Ca and Mg would, in this case, indicate a non-uniform spatial abundance structure.
           In the rightmost panels, the comparison is between the November 1 2010 spectrum and the March 4 2011 spectrum due to wavelength coverage.    
          }
\end{figure}

\subsection{Modelling}
We model the spectra obtained on June 8 2010, July 8 2010, September 5 2010, and March 4 2011 (100, 130, 190 and 370\,days after estimated explosion respectively). 
We start by modeling the blue quasi-continuum\footnote{As the spectra are calibrated to host-subtracted photometry, this component cannot be attributed to residual 
host contamination}, that seems to be the strongest component in late time spectra (March 4 and July 5 2011). 
We use the spectral continuum derived from observations of SN 2005cl, a hydrogen-rich Type IIn SN, 
with prominent blue quasi-continuum associated with CSM-ejecta interaction. It is dominated by overlapping FeII lines, some of which are potentially 
resolved \citep{Kiewe2012}. After removing the Balmer lines, and fitting a smooth curve to the continuum, we scale the resulting curve to fit the continuum seen in each 
of the SN 2010mb observed spectra. The match between the SN 2010mb blue quasi continuum, and the model based on the SN 2005cl spectrum, lends credence to this approach.\\
After removing the blue quasi continuum from the four spectra, we fit a photospheric model. SN 1997ef and SN 2010mb both show an extended LC and a late transition from 
photopheric to nebular emission (Fig. 5 and Fig. 7). We therefore use photospheric models based on SN 1997ef \citep{Mazzali2000} for SN 2010mb.   
After removing the photospheric component and following the methodology of Mazzali et al. 2004, we are left with nebular spectra.\\ 
To model the nebular component of SN 2010mb, we construct custom models, using nebular models of SN 1997ef \cite{Mazzali2000,Mazzali2004,Mazzali1993,Lucy1999,Mazzali2007} as a starting point. We assume that the emitting nebula is spherical, and has a sharp outer boundary, defined by the width of the emission lines. 
We adopt a boundary velocity of $5,000\,$km/s, in agreement with the nebular OI lines, the NaID, the CaII/[CI] blend near $8600$\,\AA, and the FeII emission near $5200$\AA, 
which is a blend of many different lines of different strength. The fact that the [CaII] lines near $7300$\AA \ and the OI $7773$\AA \ line have narrow cores, while the other 
lines do not, indicates that additional complexity beyond our simple assumption exists, see discussion. Based on the evolution of the LC and the spectra, e.g. the large 
flux fraction at late times from the blue quasi-continuum, we introduce an external energy source to the nebular model that increases with time, which we associate 
with high energy photons coming from interaction of the SN ejecta with hydrogen-free CSM.\\ 
The decomposition is shown in Fig. 7, while Table 2 summarises our results. The excellent fit between the models and the observed spectra confirms our assumption regarding 
the three components that comprise SN 2010mb observed spectra, as well as the use of SN 2005cl spectra and SN 1997ef models as starting points. 
\begin{figure}[h!p!]
\centerline{
\begin{tabular}{cc}
\scalebox{0.5}{\includegraphics{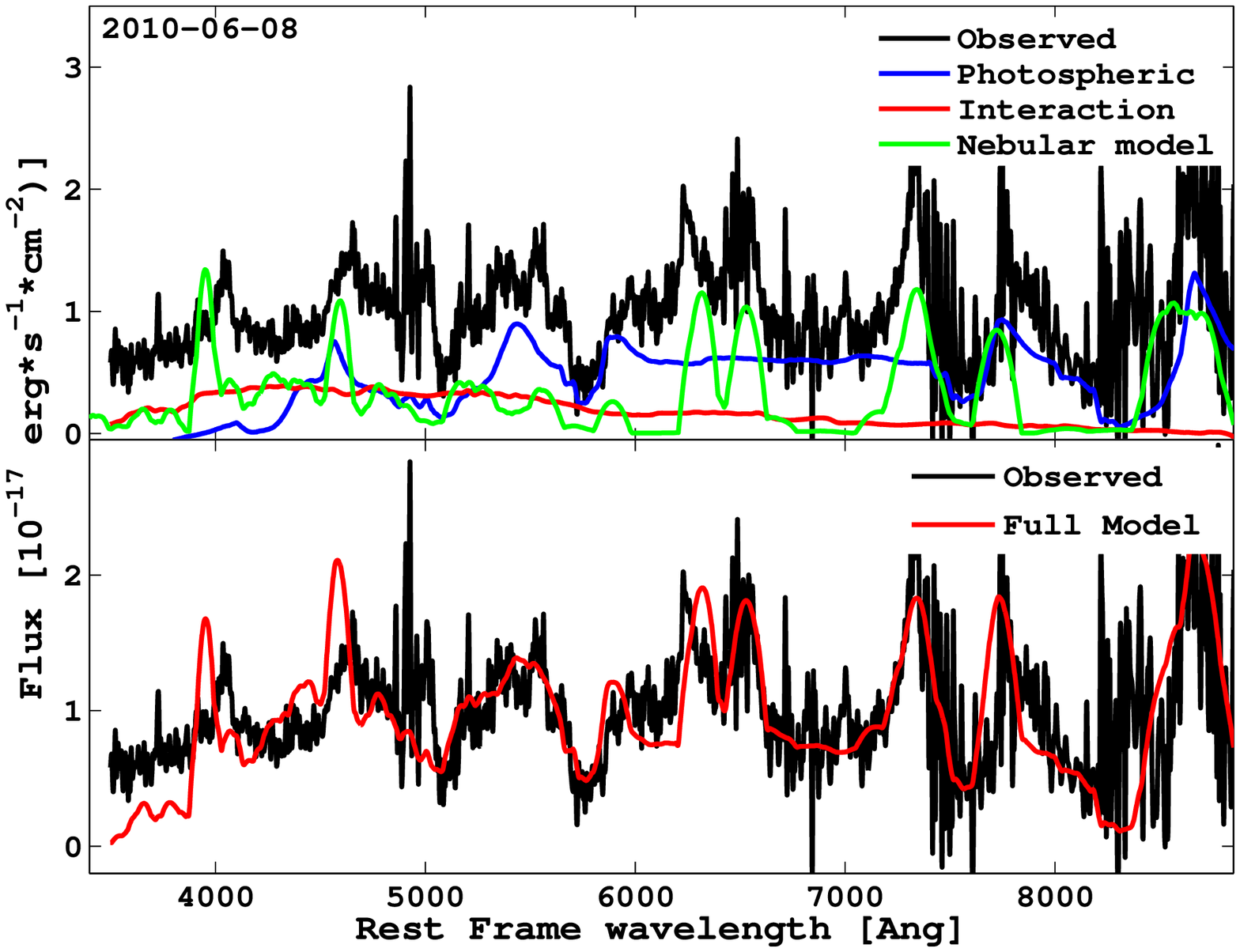}} &
\scalebox{0.5}{\includegraphics{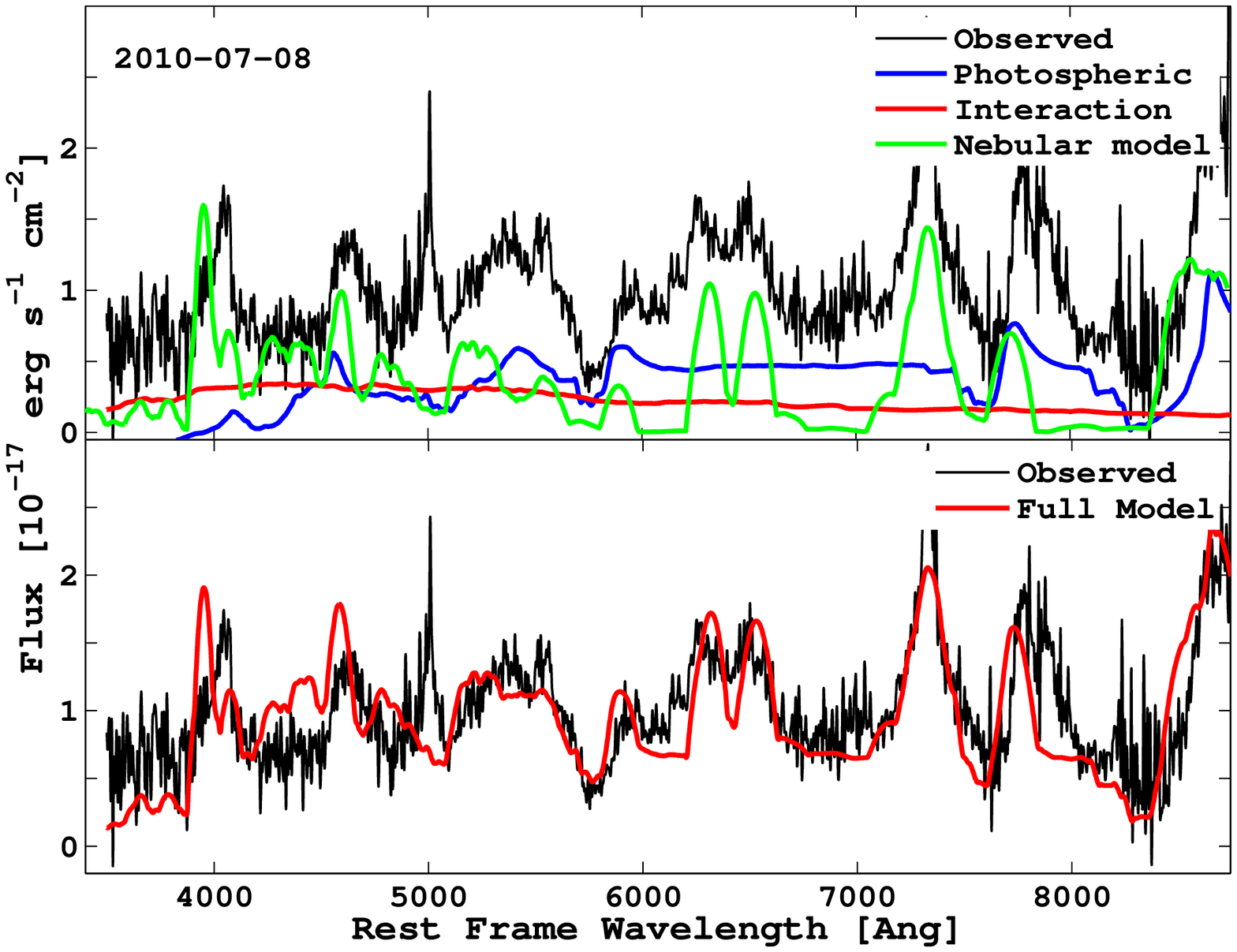}} \\
\scalebox{0.5}{\includegraphics{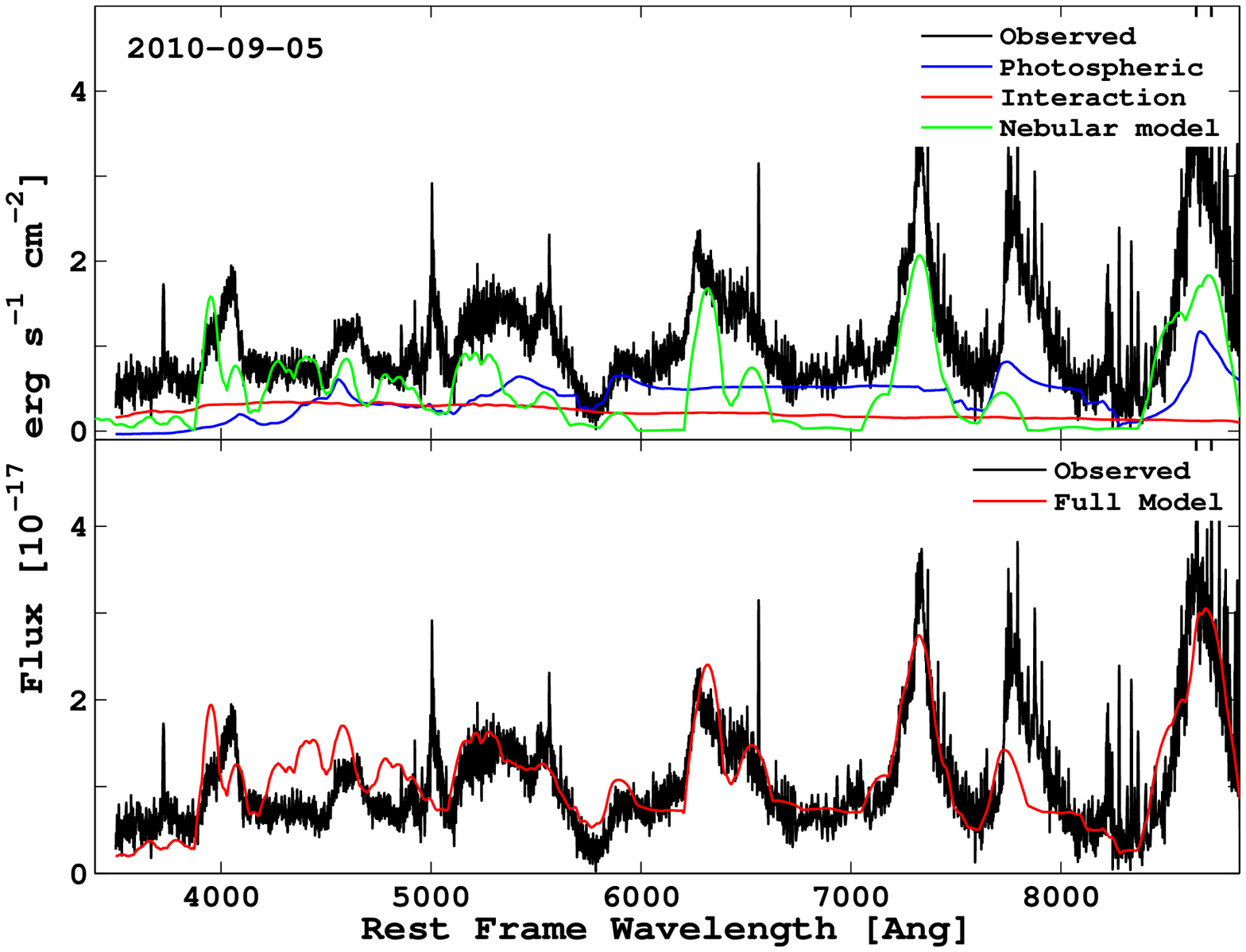}} &
\scalebox{0.5}{\includegraphics{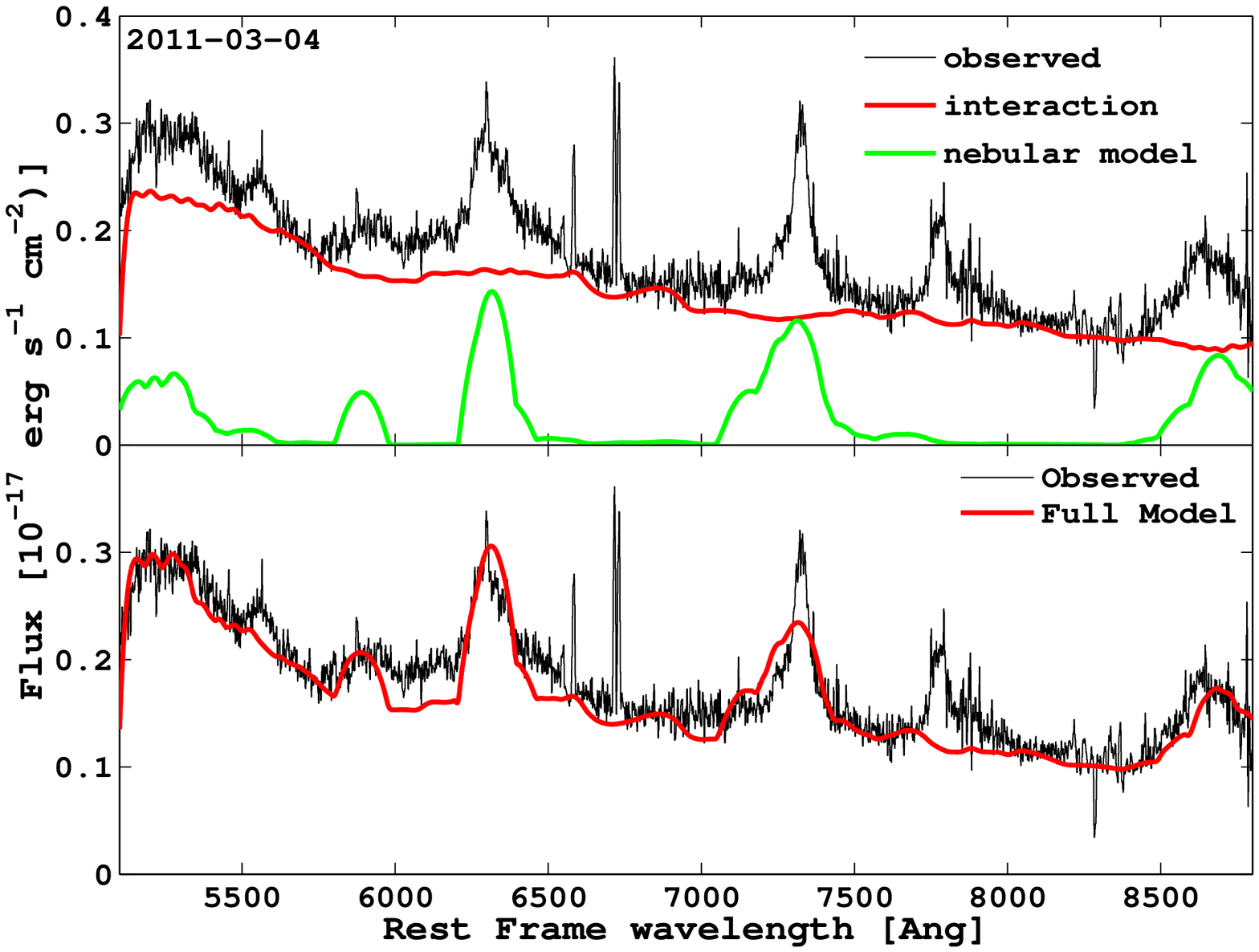}}  \\
\end{tabular}}
\caption{\small SN 2010mb Spectral Decomposition.
We decompose each spectrum to three components: photospheric and nebular models based on SN 1997ef \citep{Mazzali2000,Mazzali2004,Mazzali1993,Lucy1999,Mazzali2007}; 
and an interaction component based on a fit to the 
continuum shape (H lines excluded) of the spectrum of SN 2005cl, a hydrogen-rich Type IIn SN, with a prominent blue quasi-continuum originating from the SN ejecta 
interacting with hydrogen-rich CSM \citep{Kiewe2012}.
On the top panels, the observed data (black) is shown with its decomposition to photospheric component (blue), nebular component (green), and interaction 
component (red). On the bottom panels the three components are combined (cyan) and compared to the original data (black). The feature around $7773$\AA \ is an OI 
recombination line that is not handled by our modeling code.}
\end{figure} 

\renewcommand{\arraystretch}{1.0}
\clearpage
\begin{table}[h!p!]
\centering
\tiny
\begin{tabular}{|c|c|c|c|c|}
  \hline
  Element & June $8$, $2010$   & July $8$, $2010$   & Sep $5$, $2010$   &  March $4$, $2011$    \\ \hline
  Si      & $7M_{\odot}$       &  $6.5M_{\odot}$    &  $7M_{\odot}$     &   $8M_{\odot}$        \\  
  Ca      & $0.27M_{\odot}$    &  $0.19M_{\odot}$   &  $0.27M_{\odot}$  &   $0.065M_{\odot}$    	\\
  O       & $3M_{\odot}$       &  $2.4M_{\odot}$    &  $4.4M_{\odot}$   &		$1.7M_{\odot}$  		\\
  C       & $0.8M_{\odot}$     &  $0.7M_{\odot}$    &  $2M_{\odot}$     &	  $1.1M_{\odot}$ 	  	\\
  Na      & $0.0001M_{\odot}$  & $0.00015M_{\odot}$ &$0.00015M_{\odot}$ &  	$0.0012M_{\odot}$ 	\\
  Mg      & $0.0006M_{\odot}$  & $0.0008M_{\odot}$  &$0.0018M_{\odot}$  &  	$0.02M_{\odot}$  	\\
  S       & $3M_{\odot}$       &  $2.5M_{\odot}$    &  $2.5M_{\odot}$   &  	$1M_{\odot}$		  \\
  Ni      & $0.09M_{\odot}$    &  $0.09M_{\odot}$   &  $0.1M_{\odot}$   & 	$0.1M_{\odot}$	   \\ \hline
  Total Mass & $14.0M_{\odot}$ &  $12.4M_{\odot}$       &  $16.9M_{\odot}$  &   $12.0M_{\odot}$	    \\ \hline
\end{tabular}
 \renewcommand{\thetable}{2}
  \label{tab2}
\caption{\small Composition and ejecta mass derived for the radiating matter. The masses and composition are based on SN 1997ef nebular models with adjustments
to each element mass, so that the nebular line strengths agree with the observations.
The difference between the four spectral analysis results can be regarded as a conservative estimate of the uncertainty in the decomposition procedure 
(assuming that the emitting mass is constant in time) which is $<50\%$ in most elements. We take the total mass derived from the March 4 2011 spectrum, in which the SN is 
well into its nebular phase, as a lower limit on the ejected mass.}
\end{table} 
Our model provides an estimate of the chemical composition and ejecta mass. The masses and composition are based on SN 1997ef nebular models with adjustments
to each element mass, so that the nebular line strengths agree with the observations. We take the total mass derived from the March 4 spectrum ($12$M$_{\odot}$), 
in which the SN is well into its nebular phase, as a lower limit on the ejecta mass. For comparison, the ejecta mass in an average Type Ic SN is $1.7$M$_{\odot}$
\citep{Drout2011}. The derived mass indicates we have observed the explosion of a massive stripped-envelope star. This is further supported by the 
long transition phase from photospheric to nebular emission; delayed disappearance of photospheric (optically thick) emission indicates a large total mass \citep{Mazzali2004}. 
The simultaneous detection of photospheric/nebular SN emission and radiation from the ejecta-CSM interaction suggests either that the outer hydrogen-free material is clumpy, 
or a non-spherical geometry of the CSM (e.g. in a disk or torus). Such a geometry allows us to observe the explosion directly, in addition to ongoing interaction in the inner
circumference of the CSM structures. The mass estimates based on the nebular model should be increased, correcting for the covering fraction of the CSM.\\
\subsection{LC decomposition}
Based on our spectral decomposition, we estimate for each point in the LC the fraction of SN flux (photospheric$+$nebular), as well as that coming from the 
ejecta-CSM interaction. We thus produce two synthetic LCs - one representing the SN, and the second representing the ejecta-CSM interaction, plotted in Fig. 8\,.
The synthetic SN LC decline is in agreement with that expected from $^{56}$Ni decay, assuming full trapping of the gamma-rays, indicating a large ejecta mass. 
The amount of radioactive nickel ($0.1\pm0.01$M$_{\odot}$) suggested by the model is also consistent with the nickel mass-peak magnitude relation for 
SNe \citep{Perets2010}, predicting $\sim0.1-0.15\,$M$_{\odot}$ of $^{56}$Ni, using our photometry.
The synthetic LC for the interaction component shows a maximum $\sim220\,$days after estimated explosion and a much slower decline, and is dominating the optical display starting from 300 
days and onwards.
 \begin{figure}[h!p!]
  \centerline{
    \scalebox{0.9}{\includegraphics{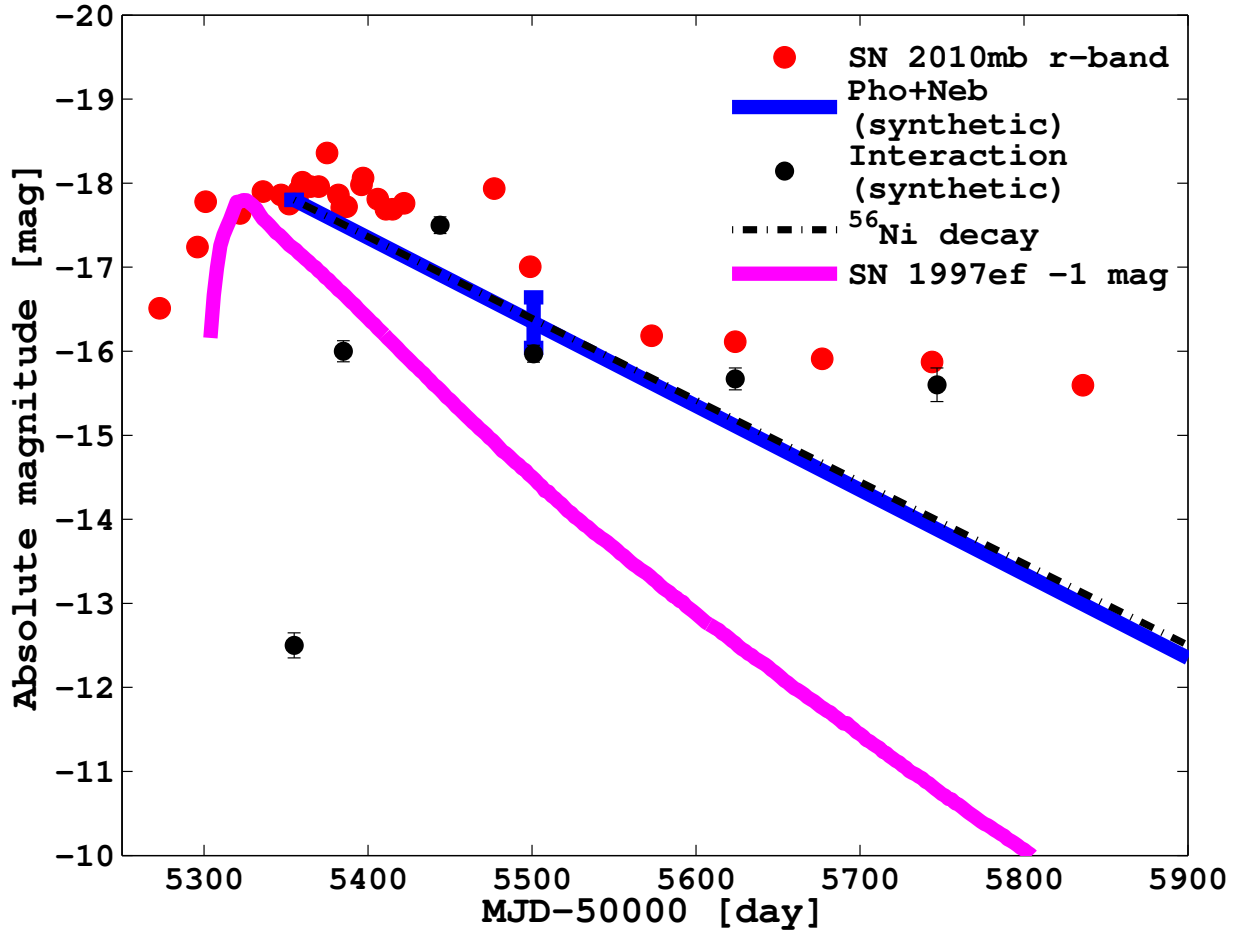}}
  }
  \caption{\small SN 2010mb synthetic LC. 
A synthetic light curve (blue), showing the decay of the photospheric and nebular components only (based on the spectral decomposition, and interpolated between spectral 
epochs) behaves as expected from a $^{56}$Ni powered SN (dashed black curve), assuming full trapping of the gamma-rays, indicating a large ejecta mass.
A synthetic LC of the interaction component is given as well (black dots), as is a scaled light curve of SN 1997ef, a slowly-declining 
Type Ic SN (magenta; Mazzali et al. 2004). The results suggest that a large fraction of the observed flux is not internal to the SN explosion, but originates from interaction
of the SN ejecta with CSM. The absence of hydrogen and helium lines in the observed spectra (Fig. 3), as well as the evolution of the narrow forbidden [OI] line at $5577$\AA{}
\ (Fig. 9), suggests it is interaction with hydrogen-free CSM.
          }
\end{figure}

\newpage
\subsection{Interaction Lines}
Some of the spectra show a prominent emission line on top of the oxygen $5577$\AA{} \ nebular feature, Fig. 9.
We detected and measured this emission line flux using the standard IRAF procedure $splot$, as well as a custom MATLAB script, which removes nearby continuum.
We verified that the line is not sensitive to varying the parameters in the nebular modeling code (i.e changing the amount of radiating mass in 
the nebular phase), and the blue quasi-continuum flux, and find a variability of less than $10\%$ for the line flux.
We ran Cloudy, a spectral synthesis code designed to predict the physical properties in the ISM under a broad range of 
conditions\footnote{Cloudy is designed to compute the ionization and thermal properties, as well as the emitted spectrum, of plasma in collisional or 
photoionization equilibrium.}  \citep{Ferland1998}. We modeled slabs of optically-thin and collisional gas, in order to compute the emission-line 
intensities associated with specific elements, as functions of the gas temperature and density. Two models are presented in Fig. 10, 
both with parameters chosen so that the flux of the [OI] line at $5577$\AA{} \ is similar to that of the lines at $6300,$ and $6363$\AA{}, \ as observed: 
(\textit{i}) A composition of S, C, O, Si, Ca, and Fe, with ratios in accordance with the March 4 2011 nebular model results 
(Table 2) in order to check whether the $5577$\AA{} \ line is coming from the SN ejecta (i.e. a slow component in the nebular matter); 
(\textit{ii}) A composition of 90\% He, 5\% C, 5\% O and traces of iron - the expected shell composition from a pulsational pair-instability (PPI) event \citep{Chatz2012}.
Comparing to the observed spectra, we rule out narrow emission from the nebular phase (which would lead to strong lines of Ca and Fe that are not seen).
The presence of only [OI] lines is consistent with the PPI model.
The property of the [OI] $5577$\AA \ line as a tracer of high densities of oxygen are shown in Fig. 10, bottom panel. Only at densities $\sim10^7\,$particles/cm$^3$ 
does the $5577$\AA \ line have a flux similar to the flux in the [OI] lines at $6300$ and $6363$\AA. For the $5577$\AA \ to be the dominant line among the [OI] forbidden 
lines as observed, a density of $\sim10^{9}$ particles/cm$^3$ is required. 
Further examination of the data shows that the $5577$\AA \ line is blueshifted by $800\,$km/s and has a velocity dispersion of $600\,$km/s, 
while the $6300$ and $6363$\AA \ lines have the same redshift as the host galaxy, and a velocity dispersion of $\sim300\,$km/s i.e, they are instrumentally unresolved.
We therefore identify the $5577$\AA \ high density line with the CSM interaction, and the $6300$ and $6363$\AA \ doublet as host-galaxy emission.\\
We find that the line at $5577$\AA \ reaches peak intensity between $\sim190$ and 
$\sim240\,$days after estimated explosion (integrated intensity of $7\pm 0.4, 7.8\pm 0.3 \cdot10^{-17}$erg/s/cm$^2$), while the $6300$ and $6363$\AA \ lines are clearly seen in the February 20 
2012 host galaxy spectrum. The observed evolution of the [OI] $5577$\AA \ line seems to be responding to the blue quasi-continuum. This behavior, as well as the resemblance of 
the blue quasi-continuum model to the spectral continuum of SN 2010mb (e.g. the March 4 2011 spectrum, Fig. 7), shows that at late times the observed LC is powered by interaction of the SN ejecta with 
hydrogen-free CSM.\\
\begin{figure}[h!p!]
\centerline{
\scalebox{0.9}{\includegraphics{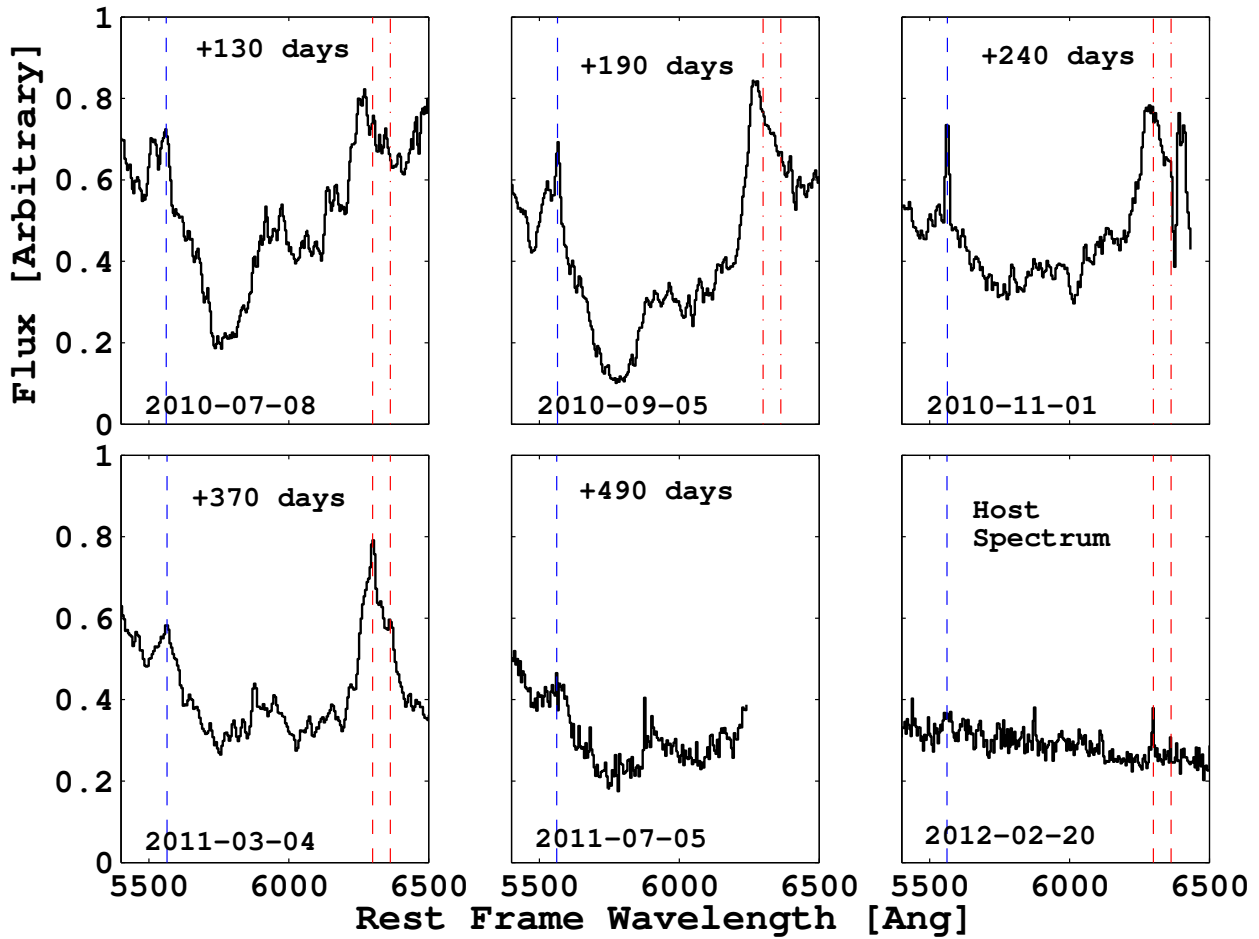}}
}
\caption{\small Evolution of the [OI] $5577$\AA \ collisionally excited  line (dashed blue), which requires high densities ($\sim10^{9}$ particles/cm$^3$ 
for the $5577$\AA \ to be the dominant among the [OI] forbidden lines), similar to those expected in a shell ejected in a PPI event.
This line is blue shifted by $\sim800\,$km/s and has a velocity dispersion of $\sim600$km/s. It is getting stronger with time between $\sim130\,$days to 
$\sim240\,$days after estimated explosion (integrated flux of $1.6\pm0.3$ and $7.8\pm0.3\times10^{-17}$\,erg/s/cm$^2$ 
$\sim130$ and $\sim240$ days after estimated explosion respectively), and traces the flux increase seen in the interaction light curve (Fig. 8). In late spectra ($\sim370$, $\sim490\,$days after 
explosion) the peak integrated flux is less than $1\times10^{-17}$\,erg/s/cm$^2$. While a PPI ejected shell is expected to be rich in He, C, and O, the [OI] $5577$\AA \ line is the only signature 
expected to fall in the optical waveband, as helium ionization requires much higher temperatures than those expected in the ejected shell, and C lines fall mainly in the near 
IR \citep{Chatz2012}. 
The [OI] lines at $6300$ and $6363$\AA \ have the same redshift as the host-galaxy and a velocity dispersion of $\sim300\,$km/s. They are clearly seen in the host galaxy 
spectrum taken on Feb 20 2012. We therefore identify them as host-galaxy emission lines (dashed red). 
}
\end{figure} 
\begin{figure}[h!p!]
\centerline{
\scalebox{0.85}{\includegraphics{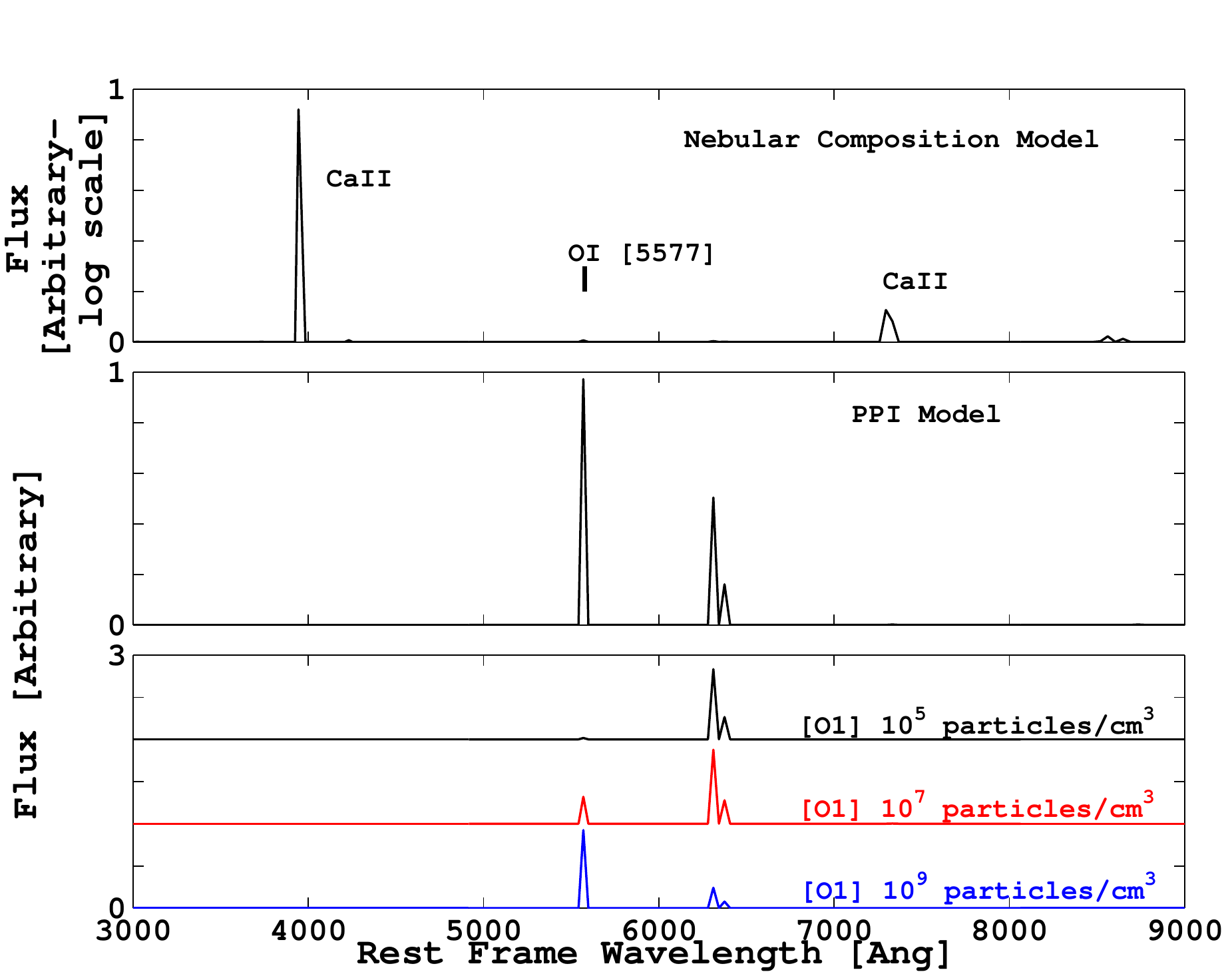}}
}
   \label{fig9}
\caption{\small SN 2010mb interaction lines.
Two models are presented, both with parameters chosen so that the flux of the [OI] line at $5577$\AA{} \ is similar to that of the lines at $6300,$ and $6363$\AA{} \ 
as observed: 
(\textit{i}) A composition of S, C, O, Si, Ca, and Fe, with ratios in accordance with the March 4 2010 nebular model results (Table 2) - top panel.
Ca lines are dominant (more than two orders of magnitude brighter than the strongest [OI] line), but are not observed. We find this model is ruled out.
(\textit{ii}) A composition of 90\% He, 5\% C, 5\% O and traces of iron - the expected shell composition from a PPI event \citep{Chatz2012} - middle panel. 
This model is in agreement with the observed lines.
The property of the [OI] $5577$\AA \ line as a tracer of high densities of oxygen is shown in the bottom panel. Only at densities $\sim10^7\,$particles/cm$^3$ (black curve), 
does the $5577$\AA \ and the $6300$ and $6363$\AA \ lines have similar fluxes. For the $5577$\AA \ to be the dominant line among the [OI] forbidden lines, 
a density of $\sim10^{9}$ particles/cm$^3$ is required (blue curve). 
}
\end{figure} 

\newpage
\subsection{CSM mass}
We can estimate the amount of CSM needed to convert the SN ejecta kinetic energy to radiation through interaction. From the light curve, we estimate the amount of 
energy radiated between consecutive data points $\Delta\mathrm{E}$, and from the line width in the observed spectra we get the ejecta velocity, v$\approx5,000\,$km/s. 
Next, using the relation $\frac{1}{2}\Delta $M$^2$v$^2=\epsilon\Delta$E, where $\epsilon$ is the conversion efficiency and is determined in an iterative way using the equation
$\epsilon=\frac{\Delta \mathrm{M}_{CSM}}{\mathrm{M}_{ej}+\Delta \mathrm{M}_{CSM}}$  \citep{Murase2011}, we derive the mass needed for energy conversion $\Delta\mathrm{M}$. 
We estimate the radius at which interaction takes place at any given time by assuming an expansion of the SN ejecta at a constant velocity of $5,000\,$km/s. 
We then derive a lower limit on the density by dividing the CSM mass by the shell volume swept by the ejecta between two consecutive points in the light curve. 
We get a total CSM mass of $\sim3.3$M$_{\odot}$ distributed around the progenitor with a density profile $\propto r^{-2.6}$, (Fig. 11, left panel). 
The derived densities ($\sim10^9\,$cm$^{-3}$ peak density) are in agreement with the observed evolution of the $5577$\AA \ [OI] recombination line.\\
Based on this analysis, we conclude the following: (\textit{i}) $\sim3$M$_{\odot}$ of hydrogen-free CSM is present at the vicinity of the progenitor, 
with mass-loss-rate higher than those associated with standard winds from WR stars, i.e. \O$(1)\,$M$_{\odot}/$year compared to \O$(10^{-5})\,$M$_{\odot}/$year. (\textit{ii}) Assuming bremsstrahlung is the main cooling mechanism of this plasma, 
the cooling time is orders of magnitude shorter than the time gap between adjacent points in the LC, i.e \O$(10^3\,s)$ at the derived densities vs. \O$(10^6\,s)\,$ between adjacent measurements on the LC, and so the CSM-ejecta 
interaction energy is radiated promptly. (\textit{iii}) The column density is high enough to explain why X-ray photons have not been detected (Fig. 11, right panel). 
\begin{figure}[h!p!]
\scalebox{0.9}{\includegraphics{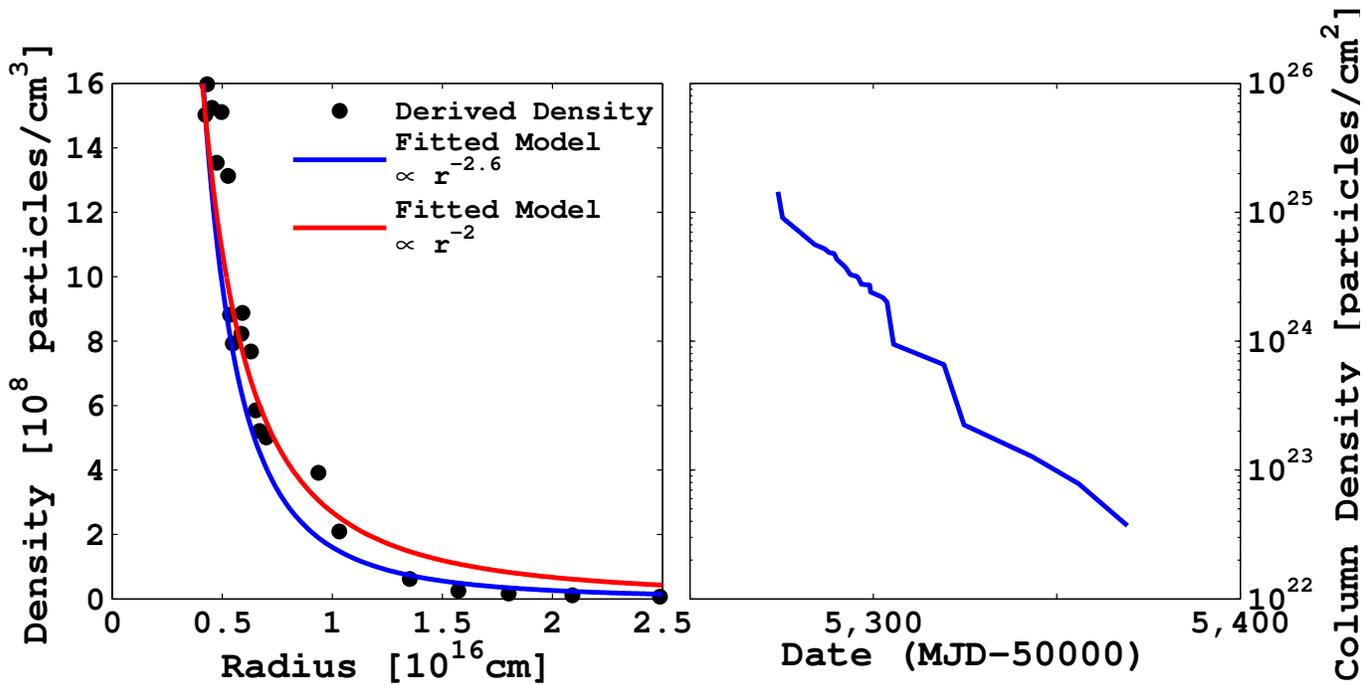}} 
\caption{Characterizing the interacting CSM.
\%textbf{Left:} A CSM density profile derived from the light curve of SN 2010mb, assuming a symmetric spherical CSM distribution (solid points; see text for details).
The data is best fitted by a power-law profile of n$\propto$r$^{-2.6}$ (blue line). For comparison, a n$\propto$r$^{-2}$ profile is also presented (red line).
\textbf{Right:} CSM column density derived by integrating over the density profile shown in the left panel. 
The high column density explains why X-ray photons have not been detected. 
}
\end{figure} 
\newpage
\subsection{Host-galaxy}
The host galaxy spectrum taken on February 20 2012 (taken at the galaxy nucleus $0.1''$ E and $2.2''$ S from the SN location; Fig. 4) shows narrow Balmer emission 
lines filling slightly wider Balmer absorption features. This is typical to galaxies with a dominant population of older stars ($\sim1\,$Gyr), mixed with a population of 
younger hotter stars \citep{Dressler1983,Covino2006}, which in turn indicates that star formation is still present in the galaxy. 
The spectrum is dominated by lines of H$\alpha$/$\beta$ ($6563$ and $4861$\AA), [NII] lines ($6548$ and $6583$\AA), [OIII] lines ($4959$ and $5007$\AA), as well as the [OII] 
line at $3727$\AA.\\  
Host galaxy analysis was performed using two different spectra: for the SN position using the spectrum taken on September 5, 2010, and for the galaxy nucleus, using 
the spectrum taken on February 20 2012. The SN has an offset of $0.1''$ W and $2.2''$ N from the host galaxy center, so a total of $2.2''$ or $5.1\,$kpc in the 
radial direction at the redshift of the host. We detected and measured the emission line fluxes of H$\alpha$, H$\beta$ [OII], [OIII] and [NII] using the standard 
IRAF procedure $splot$. We find that the host galaxy properties derived from the nucleus and from the SN position agree with each other within the error bars, with the SN 
position values having larger errors because of the lower S/N at the SN position. We estimate a reddening of E(B-V)$=0.19\pm0.03\,$mag for the nucleus and 
$E(B-V)\leq0.09$ mag for the SN site, having used the observed ratio of the Balmer lines, and assuming the Cardelli et al. (1989) extinction law with 
$R_V$=3.1, as well as case B recombination with a Balmer decrement of H$\alpha$/H$\beta$=2.86 . In the following, we correct the detected emission line fluxes for reddening.
For computing the oxygen abundance from these HII region emission lines, we follow Modjaz et al. (2011; and reference therein), and employ the scales of Pettini \& Pagel (2004) (PP04-O3N2) and of 
Kewley \& Dopita (2002) (KD02), respectively, to obtain oxygen abundance values for the host galaxy nucleus of $12+log$(O/H)$_{PP04-O3N2} = 8.39 \pm 0.01$, and 
$12+\log$(O/H)$_{KD02}=8.60\pm0.05$, respectively, where we consider only statistical uncertainties. The oxygen abundance at the SN site is consistent with that of the host 
galaxy nucleus within the error bars. We conclude that the metallicity of the SN host galaxy is  
$0.5^{+0.42}_{-0.19}$Z$_{\odot}$, having used the scale of Pettini \& Pagel (2004) and a solar oxygen abundance value of $12+\log$(O/H)$=8.69$ \citep{Asplund2009}.
The total star formation rate we measure using the H$\alpha$ luminosity (Kennicutt et al. 1998) is $\sim0.32\,$M$_{\odot}\,$yr$^{-1}\,$.\\
\newpage
\section{Discussion}
\subsection{Indications for a PPI event}
The observed data is consistently explained by a PPI event followed by a CC SN. 
During the first $\sim150\,$days after the explosion, $\sim90\%$ of the optical display is coming from the SN explosion. At later times, as the ejecta reach the 
hydrogen-free CSM, a shock wave is generated at the interface between the two media and X-ray photons generated at the shock front are exciting both the CSM and the ejecta, thus 
introducing a new source of energy. Eventually, this new source dominates the optical radiation - as can be seen from the long-lasting optical display compared to 
other Type Ic SNe, the change in the spectral energy distribution, and the evolution of the forbidden [OI] line.
If we assume a non-spherical geometry for the CSM, the high energy photons will excite mostly the ejecta in the plane of the torus, thus perhaps 
generating the observed internal structure in the nebular lines (Fig. 4)\,.\\
A recent model of a $110$M$_{\odot}$ star in a low metallicity environment produces a stripped C/O core when encountering the pair instability for the first time \citep{Chatz2012}. 
The period of pair instability will result in the ejection of a $\sim3$M$_{\odot}$ shell that is significantly enhanced in He, C, and O. 
Subsequent pulses may be even richer in carbon and oxygen, as they probe inner regions of the progenitor.
Later, the progenitor will return to the normal evolution track, until it ends its life in an iron-core-collapse SN.
In the case of SN 2010mb, we estimate the time gap between the last pulse and the core collapse event was $\sim2.2\,$years, if we assume a velocity of $800\,$km/s for the 
matter ejected in the final pulse based on the observed $5577$\AA \ [OI] recombination line blueshift, again in accordance with models \citep{Woosley2007}.\\
While in the past PISNe and PPI events were assumed to happen only in very low metallicity environments, recent works \citep{Langer2007,Chatz2012,Yusof2013} predict 
these explosions can also occur in local dwarf galaxies with metallicity values which are well within the errors of those measured for SN 2010mb, i.e. metallicities as 
high as  Z=Z$_{\odot}/3$\,.
\subsection{Comparison to Type Ibn SNe}
Interaction of SN ejecta with hydrogen-free material is also observed in Type Ibn SNe  \citep{Pastorello2008a,Smith2012}. 
These events are defined by strong HeI emission lines, and are probably associated with the explosion of Wolf-Rayet (W-R) progenitors embedded in a helium-rich hydrogen-diluted CSM. 
The interacting material composition is consistent with that of W-R winds, i.e. hydrogen and helium for WN progenitors, and helium for WC/WO progenitors  
\citep{Pastorello2008b}. 
The main differences between Type Ibn SNe and SN 2010mb are: (\textit{i}) Type Ibn SNe have much shorter time scales than that of SN 2010mb, and their B and $g$ magnitudes 
decrease with time; (\textit{ii}) While in Type Ibn the interaction has a modest influence on the observed LC, and a fairly modest amount of CSM is involved (less than 
$\sim0.3$M$_{\odot}$; Pastorello et al. 2008a), in the case of SN 2010mb, CSM interaction is the dominant power source at late times, and probably involves a much higher mass of CSM  ($>3M_{\odot}$). 
We conclude that a similar scenario to the one generating Type Ibn SNe is not adequate for SN 2010mb.\\
\subsection{Other scenarios}
\subsubsection{Photon Diffusion}
Type II-P SNe also exhibit a plateau in their LCs. The mechanism driving this plateau is photon diffusion through the expanding ejecta. We analyze the SN 2010mb LC in the 
context of the recombination front model \cite{Popov1993}. In this model, the opacity is approximated by a step function, 
and is constant above the ionization temperature, $T_{ion}$, and equals zero below that. We assume that above the ionization temperature the opacity is dominated by 
Thomson scattering, and since the envelope contains mainly metals, we take $k=0.2\,$cm$^2$g$^{-1}$. This approximation to the opacity is an upper bound on the real opacity, 
which is already smaller than $0.2$cm$^{2}$g$^{-1}$ at temperatures of $T_{ion}\approx3\times10^4$ $^{\circ}$K \cite{Rabinak2011}. We assume that the effective temperature 
of the radiation is $T_{eff}\approx T_{ion}2^{\frac{1}{4}}\approx5000^{\circ}$K. Based on this analytic model, we get the following estimation for the ejecta mass, explosion 
energy, and progenitor radius:

\[
 M_{ej}\sim 540M_{\odot}t^4_{150}T^4_{5054}v^3_{5}L_{41}^{-1}\kappa_{0.2}^{-1}
\]
\[
 E\sim3.9\times10^{52}t^4_{150}T^4_{5054}v^5_{5}L_{41}^{-1}\kappa_{0.2}^{-1}\mathrm{erg}
\]
\[
 R_s\sim5\times10^{11}t^{-2}_{185}T^{-4}_{5054}v^{-4}_{5}L_{41}^2 \kappa_{0.2}\mathrm{cm} 
\]
where $t_{150}=t_p/150$days, $T_{5054}=T_{ion}/5054^{\circ}$K, $L_{41}=L/10^{41}$erg s$^{-1}$, $v_{5}=v_{ph}/5\times10^8$cm s$^{-1}$, and $\kappa_{0.2}=\kappa/0.2$cm$^2$g$^{-1}$. 
Even when considering the uncertainties in the measurements, the model suggests an unrealistically large ejecta mass (several hundred solar masses of He+C+O), and therefore 
we pursue it no more.\\
Another event where interaction of SN ejecta with H-free CSM was suggested as a way to explain an increased, short lived luminosity is SN 2009dc, a SN interpreted as a 'Super Chandrasekhar' Type Ia SN (Hachinger et al. 2012).\\
\subsubsection{Mass loss through Wave Excitation}
The presence of a large amount of CSM can be attributed to wave excitation by vigorous convection in the late stages of stellar evolution  \citep{Quataert2012}. 
In this scenario, convective motions excite internal gravity waves that in some cases are converted to sound waves as they tunnel towards the stellar surface. 
As the sound waves dissipate while crossing through the star envelope, they release a large amount of energy that can unbind up to several solar masses of the stellar 
envelope. While this scenario allows high mass loss rates for stripped-envelope stars during Si burning, it remains to be demonstrated that high mass loss rates can be produced 
during earlier nuclear burning stages in such progenitors (carbon, neon and oxygen burning), as would be required to explain SN 2010mb.\\
\subsubsection{Magnetar}
Magnetars, very highly magnetic ($B>10^{14}\ $G) and rapidly rotating ($P_s\approx 2-5\ $ ms)
neutron stars (NSs), generated at the time of SN explosion, can have a large impact on supernova 
light curves \citep{Mazzali2006,Maeda2007,Kasen2010,Woosley2010}.  
These magnetars spin down in a few months (comparable to the radiative diffusion time in SN ejecta, $t_d$), depositing
enough energy to power the rare superluminous ($L>10^{44} \ {\rm erg
\ s^{-1}}$) hydrogen poor SLSN-I \cite{Quimby2011,GalYam2012}. The much lower 
(and nearly constant) luminosity, $L_{\rm SN} \approx
10^{42}\ {\rm erg \ s^{-1}}$, of SN 2010mb over a prolonged ($>500\,$days)
period requires a different interpretation. Specifically, a young NS
with a magnetic dipole spin-down luminosity $L_p\approx  10^{42}\ {\rm erg \ s^{-1}}$,
and a spin-down timescale, $t_p$, 
longer than 10 years, so that $L_p$ is approximately constant over the $500\,$days, is needed. Such a power source \citep{Kasen2010} would 
overwhelm 
the adiabatic losses from the initial explosion, and reset the SN luminosity to $L_p$. 
Setting $L_p=10^{42}\ {\rm erg \ s^{-1}}$ gives the relation 
$B\approx 3\times 10^{13} {\rm G}(P_s/8 \ {\rm ms})^2$, and requiring a 
spin-down timescale longer than 10 years implies that $P_s< 8 $ ms. Though more rapidly 
rotating than the fastest known young pulsar, PSR J0537-6910 \citep{Marshall1998} (at $P_s=16\,$ms), 
such an initial spin and magnetic field are certainly allowed (e.g. the implied 
$B\approx 4\times 10^{12} \ {\rm G}$ for an initial spin of 3 ms is entirely reasonable). 
In this scenario, the young NS spin-down luminosity is  reprocessed by the SN ejecta, 
allowing for a long, visible event. Though we remain unsure as to the form (i.e. Poynting flux, particles, radiation) 
of the spin-down power, $L_p$, we will assume that the thinning of the ejecta due to expansion will eventually 
lead to an inefficient coupling and hence a reduction in the optical brightness of the event. 
For an ejecta mass of $M_{\rm ej}\approx 5 M_\odot$ moving at
an average velocity fixed by $E=M_{\rm ej}V^2/2$, the column depth in the ejecta will evolve with time as 
\begin{equation} 
{M_{\rm ej} \over 4\pi (Vt)^2}= 2\  {\rm g \over cm^{2}}\left(500\  {\rm d}\over t\right)^2\left(10^{51} \ {\rm erg}\over E\right)\left(M_{\rm ej}\over 5 M_\odot\right)^2,
\end{equation}  
implying thinning to fast energetic ions and hard radiation at about the time where the SN faded.\\
The magnetar model can explain the LC timescale but fails to explain the color evolution of SN 2010mb, the high density narrow O lines, and the presence of strong nebular lines that require an optically thin medium, while the reprocessing of the radiation emitted by the magnetar requires a thick medium.

\newpage
\section{Conclusions}
Observations of SN 2010mb give direct evidence for interaction of SN ejecta with a large amount of hydrogen-free CSM. 
The extended light curve, the rise in magnitude in the $g$ and B-bands, and the blue quasi-continuum that becomes more significant with time, show that a mechanism 
external to the SN explosion is injecting energy at late times. The presence of the forbidden [OI] line at $5577$\AA \ and its evolution shows the external source 
is interaction of the SN ejecta with a large mass of hydrogen-free CSM at high densities at the vicinity of the SN. Finally, the late transition from photospheric to nebular emission
and the spectral decomposition based on the similar SN 1997ef shows that SN 2010mb ejected a large mass $\geq12$M$_{\odot}\,$, compared to $1.7$M$_{\odot}$ in average
Type Ic SNe, as is expected in an explosion of an extremely massive progenitor.\\  
Combining all theses signatures, a PPI scenario naturally comes to mind. In such a scenario, the period of PPI will result in the ejection of a few solar masses
of matter that is composed mostly of He, C, and O. After the PPI period, the progenitor will return to the normal evolution track associated with massive stars, 
until it ends its life in a CC SN.
The optical signature of such a scenario is a long-lasting SN, where the ejecta of the SN will interact with the large amount of CSM ejected in the recent past.  
Other scenarios, such as wave-driven mass loss during late stages of nuclear burning, may be possible as well. 
A similar, if more intense, hydrogen-free CSM interaction may explain the energy source behind SLSN-I events (Quimby et al. 2011, Chomiuk et al. 2011, Pastorello et al. 2010, 
Leloudas et al. 2012, see Gal-Yam 2012 for a review), which are found in growing numbers by various sky surveys in recent years.\\
\newline
The authors thank C. Franson, S.E. Woosley and K. Maeda for useful discussions on the theoretical aspects of SN 2010mb.\\
S.B.   acknowledges support by a Ramon Fellowship from ISA.\\
A.G.   acknowledges support by grants from the ISF, BSF, GIF, Minerva, the FP7/ERC grant n\textsuperscript{\underline{o}}\,307260, the "Quantum-Universe" I-core program of the planning and budgeting committee and the ISF, and a Kimmel Investigator award.\\
P.A.M. acknowledges financial support from INAF/PRIN 2011 and ASI.\\
D.P. is supported by the Alon fellowship for outstanding young researchers, and the Raymond and Beverly Sackler Chair for young scientists.
M.S.   acknowledges support from the Royal Society.\\
J.S.B. was partially supported by an NSF-CDI grant.\\
M.M.K. acknowledges generous support from the Hubble and Carnegie-Princeton Fellowships.\\
E.O.O. acknowledges the Arye Dissentshik career development chair and a grant from the Israeli MOST.\\
The National Energy Research Scientific Computing Center, supported by the Office of Science of the U.S. Department of Energy, provided staff, computational resources, and data storage for this project.\\

\end{document}